\documentclass{emulateapj}
\usepackage{apjfonts}

\begin{document}

\def\sarc{$^{\prime\prime}\!\!.$}
\def\arcsec{$^{\prime\prime}$}
\def\arcmin{$^{\prime}$}
\def\degr{$^{\circ}$}
\def\seco{$^{\rm s}\!\!.$}
\def\ls{\lower 2pt \hbox{$\;\scriptscriptstyle \buildrel<\over\sim\;$}}
\def\gs{\lower 2pt \hbox{$\;\scriptscriptstyle \buildrel>\over\sim\;$}}

\title{Self-Consistent Models of the AGN and Black Hole Populations:
   Duty Cycles, Accretion Rates, and the Mean Radiative Efficiency}

\author{Francesco Shankar\altaffilmark{1}, David H. Weinberg\altaffilmark{1}
and Jordi Miralda-Escud\'{e}\altaffilmark{2}}
\altaffiltext{1}{Astronomy Department, Ohio State University,
    Columbus, OH-43210, U.S.A.}
\altaffiltext{2}{Institut de Ci\`encies de L'Espai (IEEC-CSIC)/ICREA,
    Bellaterra, Spain}


\begin{abstract}
We construct evolutionary models of the populations of active galactic nuclei
(AGN) and supermassive
black holes, in which the black hole mass function grows at the rate
implied by the observed luminosity function, given assumptions about the
radiative efficiency and the luminosity in Eddington units. We draw on a variety of recent
X-ray and optical measurements to estimate the bolometric AGN luminosity
function and compare to X-ray background data and the independent estimate
of Hopkins et al.\ (2007) to assess remaining systematic uncertainties.
The integrated AGN emissivity closely tracks the cosmic star formation
history, suggesting that star formation and black hole growth are closely
linked at all redshifts. We discuss observational uncertainties in the
local black hole mass function, which remain substantial, with estimates of the
integrated black hole mass density $\rho_{\bullet}$ spanning the range
$3 - 5.5 \times 10^5\, {\rm M_{\odot}\, Mpc^{-3}}$. We find good agreement with
estimates of the local mass function for a reference model where all active black holes have a fixed
efficiency $\epsilon = 0.065$ and $L_{\rm bol}/L_{\rm Edd} \approx 0.4$
(shifting to $\epsilon = 0.09$, $L_{\rm bol}/L_{\rm Edd} \approx 0.9$ for
the Hopkins et al.\ luminosity function). In our reference model, the duty
cycle of $10^9 M_\odot$ black holes declines from 0.07 at $z=3$ to 0.004 at
$z=1$ and $10^{-4}$ at $z=0$. The decline is shallower for less massive
black holes, a signature of ``downsizing'' evolution in which more massive
black holes build their mass earlier. The predicted duty cycles and AGN
clustering bias in this model are in reasonable accord with observational
estimates. If the typical Eddington ratio declines at $z<2$, then the
``downsizing'' of black hole growth is less pronounced. Models with reduced
Eddington ratios
at low redshift or black hole mass predict fewer low mass black holes
($M_{\bullet}\lesssim 10^8\, {\rm M_{\odot}}$) in the local universe, while models
with black hole mergers predict more black holes at $M_{\bullet}>10^9\, {\rm M_{\odot}}$.
Matching the integrated AGN emissivity to the local
black hole mass density implies $\epsilon = 0.075 \times
(\rho_{\bullet} / 4.5\times 10^5\, {\rm M_{\odot}\, Mpc^{-3}})^{-1}$
for our standard
luminosity function estimate, or 25\% higher for Hopkins et al.'s estimate.
It is difficult to reconcile current
observations with a model in which most black holes have the high efficiencies
$\epsilon \approx 0.16-0.20$ predicted by MHD simulations of disk accretion.
We provide electronic tabulations of our bolometric luminosity function and our reference model
predictions for black hole mass functions and duty cycles as a function of
redshift.
\end{abstract}

\keywords{cosmology: theory -- black hole: formation -- galaxies:
evolution -- quasars: general}

\section{Introduction}
\label{sec|intro}

The long-standing hypothesis that quasars are powered by accretion
onto supermassive black holes (Salpeter 1964; Lynden-Bell 1969; Rees
1984) is now strongly supported by many lines of evidence, including
the apparent ubiquity of remnant black holes in the spheroids of
local galaxies (Richstone et al. 1998). The strong correlations
between the masses of central black holes and the luminosities,
dynamical masses, and velocity dispersions of their host spheroids
(e.g., Magorrian et al. 1998; Ferrarese \& Merritt 2000; McLure \&
Dunlop 2002; Marconi \& Hunt 2003; H\"{a}ring \& Rix 2004; Ferrarese
\& Ford 2005; Greene \& Ho 2006; Graham 2007; Hopkins et al. 2007b;
Shankar \& Ferrarese 2008; Shankar et al. 2008) imply that the
processes of black hole growth and bulge formation are intimately
linked. Theoretical models typically tie black hole growth to
episodes of rapid star formation, perhaps associated with galaxy
mergers, and ascribe the black hole-bulge correlations to energy or
momentum feedback from the black hole or to regulation of black hole
growth by the bulge potential (e.g., Silk \& Rees 1998; Kauffmann \&
Haehnelt 2000; Cavaliere \& Vittorini 2000; Granato et al. 2004,
2006; Murray, Quataert, \& Thompson 2004; Cattaneo et al. 2005;
Miralda-Escud\`{e} \& Kollmeier 2005; Monaco \& Fontanot 2005;
Croton et al. 2006; Hopkins et al. 2006a, Malbon et al. 2006). These
correlations also make it possible to estimate the mass function of
black holes in the local universe (e.g., Salucci et al. 1999; Yu \&
Tremaine 2002; Marconi et al. 2004; Shankar et al. 2004, S04
hereafter). This local mass function provides important constraints
on the co-evolution of the quasar and black hole populations. The
most general and most well known of these constraints is the link
between the integrated emissivity of the quasar population, the
integrated mass density of remnant black holes, and the average
radiative efficiency of black hole accretion (So{\l}tan 1982; Fabian
\& Iwasawa 1999; Elvis et al. 2002).

In the paper we construct self-consistent models of the quasar population
using a method that can be considered a ``differential'' generalization
of So{\l}tan's (1982) argument.
Given assumed values of the radiative efficiency and the Eddington ratio
$L/L_{\rm Edd}$, the observed luminosity function of quasars at a given redshift can be
linked to the average growth rate of black holes of the corresponding mass,
and these growth rates can be integrated forward in time to track the evolution
of the black hole mass function. This modeling approach has been developed and applied
in a variety of forms by numerous authors, drawing on steadily improving observational data
(e.g. Cavaliere et al. 1973; Small \&
Blandford 1992; Salucci et al. 1999;
Cavaliere \& Vittorini 2000; Yu \& Tremaine 2002;
Steed \& Weinberg 2003, hereafter SW03; Hosokawa 2004; Yu \& Lu 2004; Marconi et al.
2004; S04; Merloni 2004; Vittorini,
Shankar \& Cavaliere 2005; Tamura et al. 2006; Hopkins et al. 2007).
The consensus of recent studies is that evolutionary models with radiative efficiencies
of roughly 10\% and mildly sub-Eddington accretion rates yield a reasonable
match to the observational estimates of the remnant mass function. However, uncertainties
in the bolometric luminosity function of active galactic nuclei (AGN, a term we will use to describe both quasars and less
luminous systems powered by black hole accretion) and in the local black hole mass function remain
an important source of uncertainty in these conclusions.

We begin our investigation by constructing an estimate of the bolometric AGN luminosity function.
Our estimate starts from the model of Ueda et al. (2003, hereafter
U03), based
on data from several X-ray surveys, but we adjust its parameters based on more recent
optical and X-ray data that provide more complete coverage of luminosity and redshift. Comparison to the X-ray background
and to the independent luminosity function estimate of Hopkins, Richards \& Hernquist (2007; HRH07 hereafter)
gives an indication of the remaining uncertainties, associated mainly with bolometric corrections and the fraction
of obscured sources. In agreement with Marconi et al. (2004) and Merloni (2004), we find
similar trends in the evolution of AGN emissivity and the cosmic star formation rate, supporting
models in which the growth of black holes occurs together with the formation of stars in their
hosts (e.g., Sanders et al. 1988; Granato et al. 2004; Hopkins et al. 2006a). We also reassess current
estimates of the local black hole mass function, finding that different choices and calibrations of
the black hole-bulge correlation lead to significantly different results, with integrated mass densities
that span nearly a factor of two.

Our simplest models of the evolving black hole and AGN populations assume that all active black holes have a single radiative
efficiency $\epsilon$ and a single accretion rate $\dot{m}$ in Eddington units. The work of Kollmeier et al. (2006)
suggests that a constant value of $\dot{m}$ may be a reasonable approximation
for luminous quasars, but the observational evidence
on this point is mixed (e.g., McLure \& Dunlop 2004;
Vestergaard 2004; Babic et al. 2007; Bundy et al. 2007; Netzer \& Trakhtenbrot 2007; Netzer et al. 2007; Rovilos \& Gorgantopoulos 2007;
Shen et al. 2007b),
and for lower luminosity, local
AGN there is clearly a broad range of Eddington ratios (e.g., Heckmann et al. 2004).
We also consider models in which $\dot{m}$ depends on redshift or black hole mass, and we consider
a simple model of black hole mergers to assess their potential impact on the mass function.
We will examine models with distributions of $\dot{m}$ values and a more realistic treatment
of mergers in future work.

Our modeling allows a consistency test of the basic scenario in which the observed luminosity
of black hole accretion drives the growth of the underlying black hole population, and
comparison to the local black hole mass function yields constraints on the average
radiative efficiency and the typical accretion rate. For specified $\epsilon$ and $\dot{m}$,
the model predicts the black hole mass function as a function of time, and combination
with the observed luminosity function yields the duty cycle as a function of redshift and black hole mass.
These predictions can be tested against observations of AGN host galaxy properties and AGN clustering,
and they can be used as inputs for further modeling. While significant uncertainties
remain, we find that a simple model of the black hole and AGN populations achieves
a good match to a wide range of observational data.

\section{The AGN bolometric luminosity function}
\label{sec|AGNLF}
In order to constrain the
accretion history of black holes, it is essential to know the shape
and evolution with redshift of the AGN bolometric LF. The
determination of this LF is not straightforward, as it must be made
using the AGN LF \emph{observed} in particular bands and converted
into bolometric estimates using empirically calibrated bolometric
corrections. The bolometric corrections have been shown by several
authors to be redshift independent, with some possible trend with
the intrinsic luminosity of the source (e.g., Elvis et al. 1994;
Marconi et al. 2004; Richards et al. 2006; Hopkins et al. 2007a). AGN
surveys have been carried out at several energy ranges, and in each
one the detection of sources can be highly affected by intrinsic or instrumental selection effects.
In different ranges of
redshift and luminosity, the best statistics come from
different wavebands or surveys, further complicating the effort to create a comprehensive
AGN LF.

An additional challenge in assembling a reliable and complete census
of the AGN population is obscuration of the central engine by gas
and dust, which may reside in the ``molecular torus'' of unified
models (Antonucci 1993) or in the interstellar medium of the
galactic host (Martinez-Sansigr\'{e} et al. 2005; Rigby et al.
2006). Strong obscuration can eliminate sources from surveys
entirely, while uncorrected weak obscuration causes their
luminosities to be underestimated. For solar composition gas, the
absorption optical depth to X-ray photons is
$\tau=2.04(N_H/10^{22}\, {\rm cm^{-2}})(E/1\, {\rm keV})^{-2.4}$
(e.g., Kembhavi \& Narlikar 1999). For Galactic dust-to-gas ratio,
the extinction in the rest-frame visual band is $A_V=5.35{\rm mag}
\times \,(N_H/10^{22}\, {\rm cm^{-2}})$ (e.g., Binney \& Merrifield
2000). Hard X-ray selection is thus the least affected by
obscuration, and deep X-ray surveys indeed reveal many faint AGNs
that are missed by traditional optical selection criteria (e.g.,
Hasinger et al. 2005). Numerous studies suggest that the incidence
of obscuration decreases towards high intrinsic luminosity, so that
X-ray and optical LFs agree at the bright end but diverge at low
luminosities (e.g., U03; Richards et al. 2005; La Franca et al.
2005; Hasinger et al. 2005). Statistically speaking, there is fairly
good correspondence between X-ray AGNs with $\log N_H/{\rm cm^{-2}}
\le 22$ and broad-line optical AGNs (e.g., Tozzi et al. 2006). Even
2-10 keV selection becomes highly incomplete for Compton-thick
sources, with $\log N_H/{\rm cm^{-2}} \ge {24}$, and the best
constraints on this population come from the normalization and
spectral shape of the X-ray background.

Given these considerations, we have chosen to base our bolometric LF
estimate principally on the work of U03, who compiled a vast sample
from \emph{Chandra}, \emph{ASCA}, and \emph{HEAO}-1 surveys and inferred
the absorption-corrected LF in the rest-frame 2-10 keV band out to $z\sim 3$.
We adopt the luminosity-dependent bolometric correction of Marconi
et al. (2004); we include a dispersion of 0.2 dex in this correction
(see, e.g., Elvis et al. 1994), but omitting the scatter would not
significantly affect our results. U03 fit their data with a parameterized
model of luminosity-dependent density evolution. We adopt the
model and adjust its parameters in some redshift ranges to better fit
other data sets, especially at high redshifts and at bright
luminosities at low redshifts.
We note that many other studies have reported
general trends of the LF evolution and obscuring columns similar to
those of U03 (e.g., La Franca et al. 2005), and that infrared surveys also
suggest a substantial population of obscured AGNs at low
luminosities (Treister et al. 2006; see also Ballantyne and Papovich
2007).

\begin{figure}
\includegraphics[angle=00,scale=0.95]{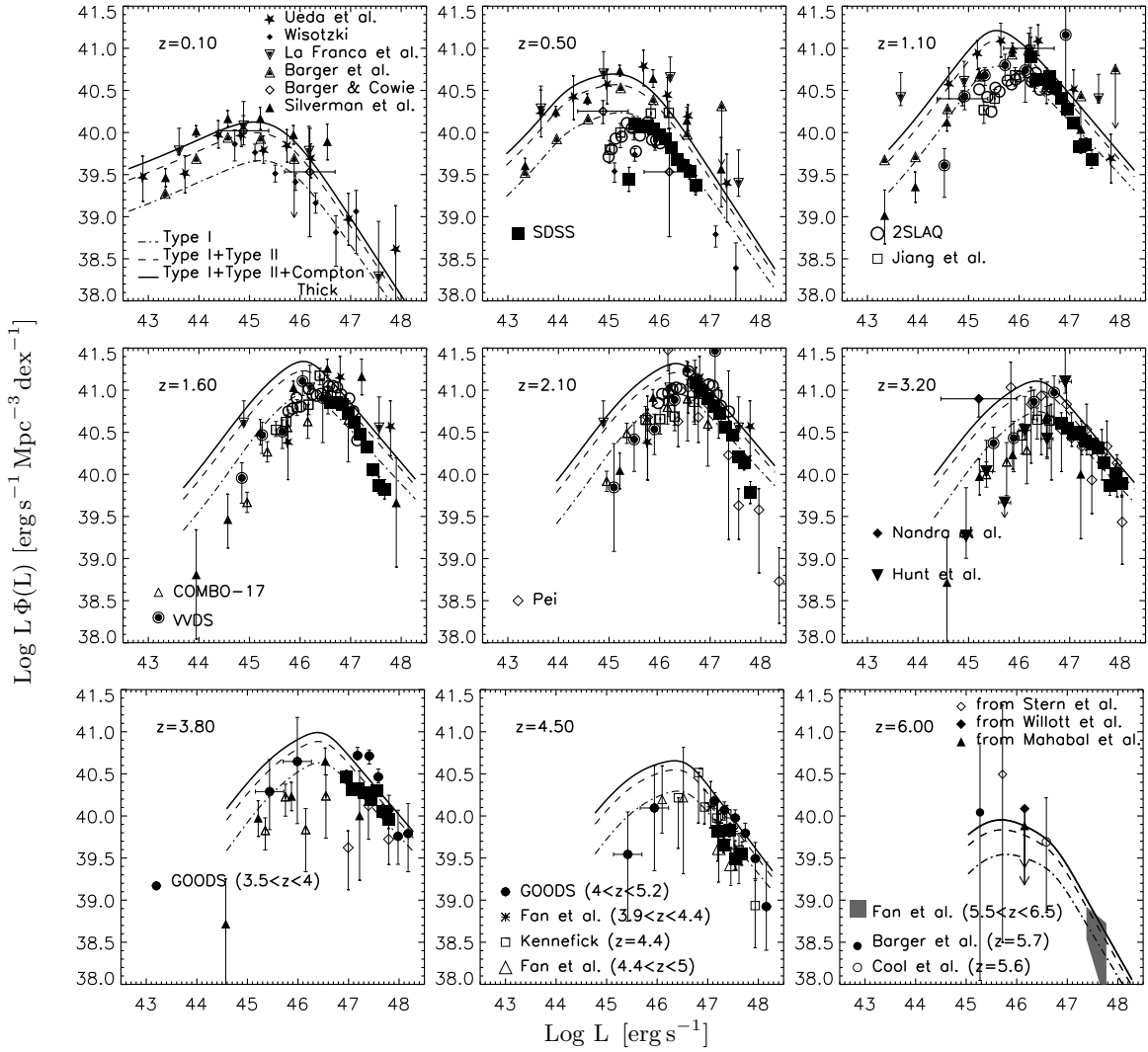}
\caption{The bolometric AGN luminosity function. Curves show the
model described in \S~\ref{sec|AGNLF}. The \emph{solid} line is the
total LF including very Compton-thick sources, the
\emph{short-dashed} line includes sources with column densities up
to $\log N_H/{\rm cm^{-2}}=24$, while the \emph{dot-dashed} line
includes only sources with $\log N_H/{\rm cm^{-2}} \le 22$. The data
from optical surveys are from Hunt et al. (2004), Pei (1995),
Wisotzki (1999), Jiang et al. (2006), Cool et al. (2006), Shankar \&
Mathur (2007; derived from data of Stern et al. 2000; Willott et al.
2004; Mahabal et al. 2005), VIMOS-VLT Deep Survey (Bongiorno et al.
2007), 2SLAQ (Richards et al. 2005), SDSS (Richards et al. 2006; Fan
et al. 2001, 2004), COMBO-17 (Wolf et al. 2003), Kennefick et al.
(1994). The data from X-ray surveys are from Ueda et al. (2003),
Barger et al. (2003), Barger et al. (2005), Barger \& Cowie (2005),
Nandra et al. (2005) and Silverman et al. (2008). GOODS
(multi-wavelength) data are from Fontanot et al. (2007). }
\label{fig|BolLF}
\end{figure}

The number of sources per unit volume per dex of luminosity $\log
L_X=\log L_{2-10\, {\rm keV}}$ is fitted by U03 using a double
power-law multiplied by an ``evolution'' term:
\begin{equation}
\Phi(L_X,z)=e(z,L_X)\frac{A}{\left(\frac{L_X}{L^*}\right)^{\gamma_1}+\left(\frac{L_X}{L^*}\right)^{\gamma_2}}\,
,
    \label{eq|U03}
\end{equation}
where
\begin{equation}
e(L_X,z)=\left\{
  \begin{array}{ll}
    (1+z)^{p_1}                      & \hbox{if $z<z_c(L_X)$} \\
    (1+z_c)^{p_1}[(1+z)/(1+z_c(L_X))]^{p_2} & \hbox{if $z\ge z_c(L_X)$\, ,}
  \end{array}
\right.
    \label{eq|eLz}
\end{equation}
with
\begin{equation}
z_c(L_X)=\left\{
  \begin{array}{ll}
    z_c^*                      & \hbox{if $L_X\ge L_a$} \\
    z_c^*(L_X/L_a)^{0.335}    & \hbox{if $L_X< L_a$\, .}
  \end{array}
\right.
    \label{eq|zc}
\end{equation}
We tune the value of parameters in order to provide a good fit to
the overall set of data presented in Figure~\ref{fig|BolLF}. The
full list of parameters is given in Table~\ref{Table|LF}. Some of
the parameters assume different values in different redshift
intervals, in which case we apply a linear interpolation in $z$ between
these intervals, as reported in
Table~\ref{Table|LF}. 
The X-ray LF of equation~(\ref{eq|U03}) is converted into a bolometric
LF using the fit to the bolometric correction given by Marconi et
al. (2004), $\log L/L_X=1.54+0.24\zeta+0.012\zeta^2-0.0015\zeta^3$
with $\zeta=\log L/L_{\odot}-12$, and $L_{\odot}=4\times 10^{33}\
{\rm erg\, s^{-1}}$, then convolved with a Gaussian scatter of
dispersion $0.2\, {\rm dex}$ to account for dispersion in the bolometric
correction.
We find the bright-end slope $\gamma_2$ should vary with time
to match the data, steepening at
very low redshifts ($z\le 0.8$) and at high ($z\ge 4$) redshifts.
We caution that the ``evolution'' term $e(L_X,z)$ substantially modifies
the shape of the luminosity function below $L\approx 10^{45}\, {\rm erg\, s^{-1}}$.
For example, the effective LF slope in the range $10^{43}\, {\rm erg\, s^{-1}}\le L\le 10^{44.5}\, {\rm erg\, s^{-1}}$
changes from $-1.4$ at $z\sim 2$ to $-0.7$
at $z\sim0$ even though we keep $\gamma_1$
(appearing in equation~[\ref{eq|U03}]) fixed at 0.86.

Figure~\ref{fig|BolLF} compares our model of the AGN LF to a large
collection of data from optical surveys (Pei 1995; Wisotzki 1999;
Fan et al. 2001, 2004; Kennefick et al. 1994; Hunt et al. 2004; Wolf
et al. 2003; Richards et al. 2005, 2006; Jiang et al. 2006; Cool et
al. 2006; Bongiorno et al. 2007; Fontanot et al. 2007; Shankar \&
Mathur 2007) and X-ray surveys (Barger et al. 2003; Ueda et al.
2003; Barger \& Cowie 2005; Barger et al. 2005; La Franca et al.
2005; Nandra et al. 2005; Silverman et al. 2008). Optical data have
been converted from the $B$-band into bolometric quantities using
$L_{\nu_B}\nu_B C_B=L$, where $L_{\nu_B}$ is the monochromatic
luminosity at $\nu_B=6.8\times 10^{14}\, {\rm Hz}$ ($\sim 4400\,$
{\AA}). We use an average value of $C_B=10.4$ consistent with
Richards et al. (2006). Note that Marconi et al. (2004) and Hopkins
et al. (2007; HRH07 hereafter) suggest values 30\%-50\% lower
depending on luminosity, while Elvis et al. (1994) proposed a higher
value of 11.8.
In Figure~\ref{fig|BolLF} we plot $L\Phi(L)$, instead of
simply $\Phi(L)$ itself, to highlight the luminosity bins where most of the
energy is emitted. For the same reason, we will later plot the black hole
mass function as $M_{\bullet}\Phi(M_{\bullet})$ to highlight the
mass bins in which most of the density resides.

The dot-dashed curves in Figure~\ref{fig|BolLF} show our model LF
including only sources with $\log N_H/{\rm cm^{-2}}\le 22$, which we
calculate using the luminosity-dependent column density distribution
functions of U03. The dashed curves show the contribution of all
sources with $\log N_H/{\rm cm^{-2}}\le 24$, while the solid curves
include higher column density (Compton-thick) systems. We assume that the number of Compton-thick
sources in each luminosity bin is equal to the number of sources in the column density
range $23\le \log N_H/{\rm cm^{-2}}\le 24$. We will
henceforth follow the X-ray convention of describing systems with
$\log N_H/{\rm cm^{-2}} < 22$ as ``Type I'' AGN and systems with
$22\le \log N_H/{\rm cm^{-2}} \le 24$ as Type II AGN. We roughly
expect the ``Type I'' curve (dot-dashed) to agree with optical survey
data points and the ``Type I+Type II'' curve to agree with X-ray
survey data points. The agreement between the model and the data is
overall fairly good, though in some regimes different data sets give
seemingly incompatible results, making it difficult to judge the
validity of the model itself. 

\begin{figure}
\includegraphics[angle=00,scale=0.95]{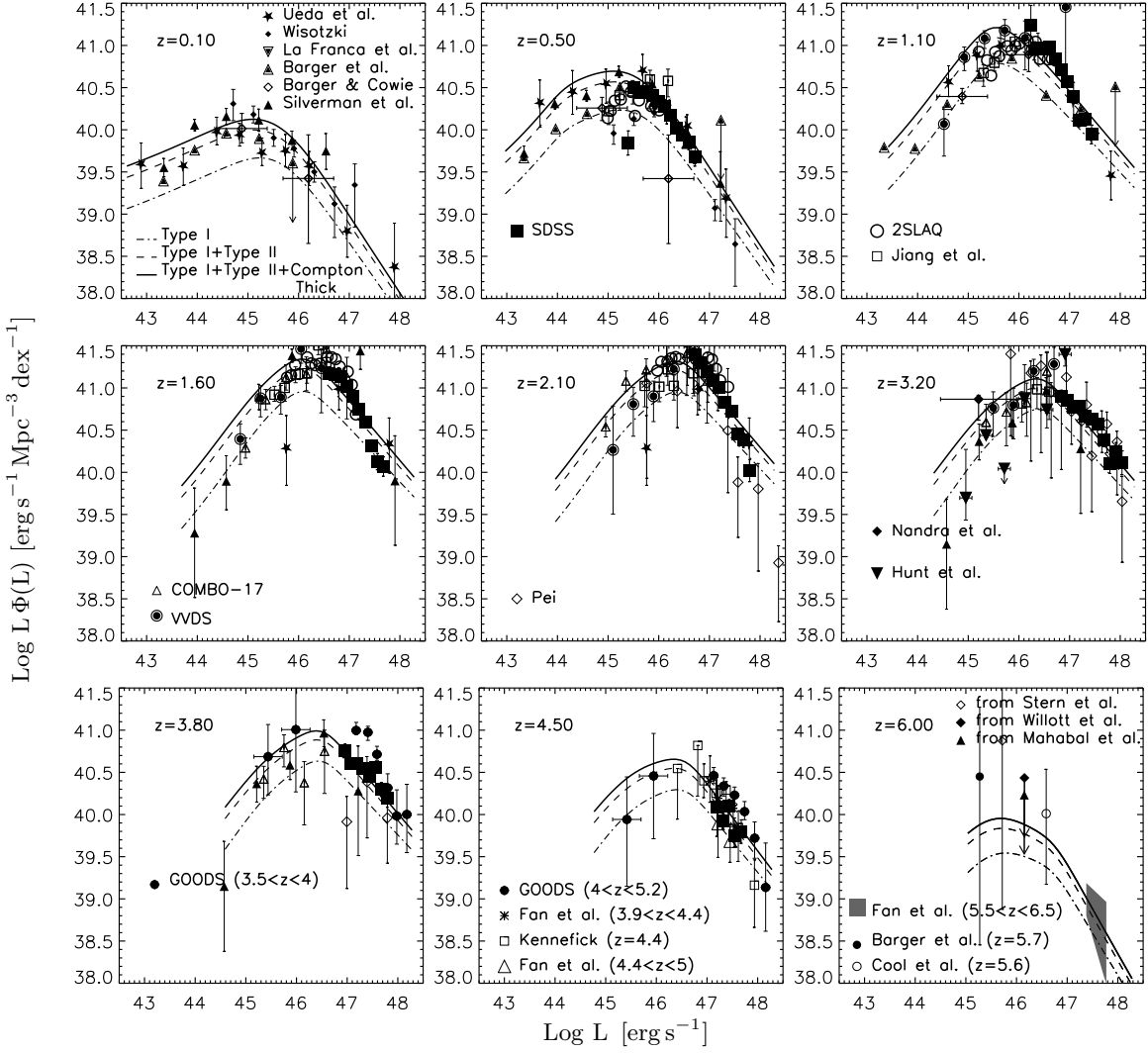}
\caption{Same as Figure~\ref{fig|BolLF} but now all the data have
been corrected for the obscured fraction as given in HRH07.} \label{fig|BolLFcorrected}
\end{figure}

HRH07 have used their observationally constrained theoretical models to estimate
the fraction of AGN that are ``missing'' from observational samples because of obscuration, as a function
of waveband and luminosity. In Figure~\ref{fig|BolLFcorrected}, we plot data points corrected
for these obscured fractions, with the same model curves as Figure~\ref{fig|BolLF}.
If the data, the obscuration corrections,
and our model were all perfect, then all of the data points should line up with each other and with the solid
model curves.
The obscuration corrections do reduce the discrepancies among data sets, most notably in the
range $z\approx 0.7-3.2$, but they do not remove all of the differences. The upper envelope of the data points
is generally close to our model near the peak of $L\Phi(L)$, though at the highest luminosities the model
exceeds the data (principally SDSS) for $z\approx 0.7-2.0$.

\begin{figure}
\includegraphics[angle=00,scale=0.9]{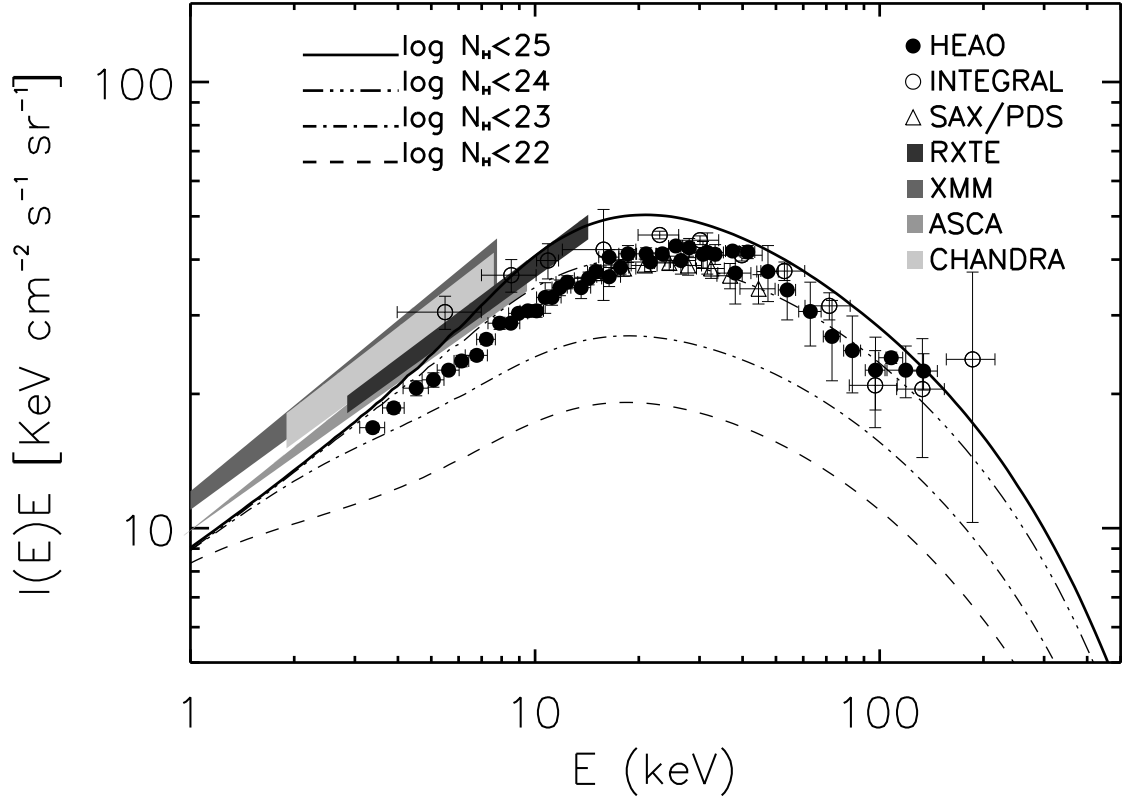}
\caption{The X-ray background. Curves represent the integration of
our AGN bolometric LF converted to hard X-ray bands, as described in
the text. The \emph{dashed} line includes sources with column
densities up to $\log N_H/{\rm cm^{-2}}\le 22$, the
\emph{dot-dashed} line those up to $\log N_H/{\rm cm^{-2}} \le 23$,
and the \emph{triple-dot dashed} line those with $\log N_H/{\rm
cm^{-2}}\le 24$. The \emph{solid} line refers to the total when
including Compton-thick sources up to $\log N_H/{\rm cm^{-2}}\le
25$, with a dependence on luminosity following that in U03. Data are
taken from a compilation in Frontera et al. (2007).}
\label{fig|XRBGs}
\end{figure}

In Figure~\ref{fig|XRBGs} we show the integrated intensity obtained
from our model AGN LF compared with all the available data on the
cosmic X-ray Background (XRBG) for energies above 1 keV. The data
are a collection of old and new results presented by Frontera et al.
(2007), plus the recent results from \emph{INTEGRAL} (Churazov et
al. 2006). While the shape of the XRBG is now well established at
low energies (2-10 keV band), different studies imply normalizations
that differ by up to 40\%. The minimum estimate for the
normalization was found by the very first \emph{HEAO} experiments
(e.g., Marshall et al. 1980; filled circles in
Figure~\ref{fig|XRBGs}), while the much more recent \emph{Chandra}
and \emph{XMM} results point towards higher values (shaded bands in
Figure~\ref{fig|XRBGs}; see Comastri 2004 for a review). Several
groups have assumed that the \emph{HEAO} measurements were correct
in shape but underestimated in normalization because of flux
calibration and/or instrumental background subtraction (Frontera et
al. 2007), implying a true XRBG spectrum that is $\sim 1.4$ times
the \emph{HEAO} spectrum at all energies. However, recent
measurements from \emph{INTEGRAL} (Churazov et al. 2006; open
circles in Figure~\ref{fig|XRBGs}) and from the \emph{PDS}
instrument aboard \emph{BeppoSAX} (Frontera et al. 2007) produce
results close to the original \emph{HEAO} measurements in the $\sim$
20-100 keV range. Based on the agreement of multiple experiments in
overlapping energy ranges, we tentatively conclude that the true
XRBG spectrum approximately follows the \emph{INTEGRAL} points over
the range $\sim$ 5-100 keV.

The solid curve in Figure~\ref{fig|XRBGs} shows the XRBG predicted
by our AGN bolometric LF model with the U03 column density
distribution. Following U03, we use the PEXRAV code (Magdziarz \&
Zdziarski 1995) to generate spectra of families of AGN with
different column densities, also taking into account Compton
down-scattering for highly obscured sources (Wilman \& Fabian
1999). We compute the spectra of Compton-thick ($\log N_H/{\rm cm^{-2}}>
24$) sources assuming $\log N_H/{\rm cm^{-2}}=24.5$. Dashed, dot-dashed, and triple-dot-dashed curves show the
cumulative contributions of sources with $\log N_H/{\rm cm^{-2}}<22,
23$, and 24, respectively. The Compton-thick sources increase the predicted XRBG by a factor of
about 1.3 at $E\gtrsim 20$ keV, with smaller contributions at lower
energies and at much higher energies. Our full model would be a good
match to the \emph{HEAO} spectrum renormalized by a constant factor
of $\sim 1.4$ as proposed in some earlier studies. However, it significantly overpredicts the
\emph{INTEGRAL} and \emph{SAX/PDS} results at $E> 10$ keV; these are
well matched by the contribution of $\log N_H/{\rm cm^{-2}}\le 24$
sources alone, with \emph{no} Compton-thick contribution. As noted
above, we employ U03's prescription for the luminosity dependence of
the Compton-thick fraction. If we use a simpler model in which
the number of Compton-thick sources is $40\%$ of the number of $\log
N_H/{\rm cm^{-2}}\le 24$ sources at all luminosities, then we obtain
very similar XRBG predictions (higher by a few percent near the peak
and even closer at other energies).

We conclude from
Figure~\ref{fig|XRBGs} that the Compton-thick fraction in our
standard model is probably an upper limit on the true fraction, and
in our calculations below we will also consider a model in which the
Compton-thick fraction is zero. However, Gilli et al. (2007) have
proposed a successful model to fit the $INTEGRAL$ measurements of the XRBG
that uses a combined contribution from
unobscured sources (which they define to be only those with $\log N_H/{\rm cm^{-2}}\le 21$)
a factor $\sim$ 2 below
our estimate and a higher contribution from obscured sources up to $\log
N_H/{\rm cm^{-2}}=26$, which they derive by assuming that these sources have a Gaussian distribution
of spectral indices around a mean slope $\langle \Gamma \rangle=1.9$.
With this result in mind, we will also explore models with a higher fraction of Compton-thick
sources (but our standard estimate of unobscured sources).

Reproducing the high background normalization
found by \emph{XMM} in the 2-5 keV range would require a
substantially higher contribution from Type I AGN, since obscured
AGN have little emission at these energies. Such an increase seems
implausible on other grounds. Following Schirber \& Bullock (2003)
and Miralda-Escud\'{e} (2003), we have estimated the average
hydrogen ionization rate produced by Type I X-ray AGN with optical
luminosity $M_B\le -15$ assuming UV spectral energy index $\alpha_{\rm
UV}=1.57$. Using the redshift-dependent X-ray-to-optical ratio by
Steffen et al. (2006), we find that the Type I AGNs produce an
emissivity that saturates the estimated UV background intensity at
$z\lesssim 1$. We therefore conclude
that the true AGN contribution to the XRBG is probably closer to our model prediction than to the $XMM$
band shown in Figure~\ref{fig|XRBGs}. If future missions confirm a high normalization of the XRBG in the 2-5 keV range,
it would probably mean that a substantial contribution from X-ray binaries in normal galaxies is present at these
energies. On the other hand, we note from Figure~\ref{fig|BolLF} that our
AGN LF defined for sources with $\log N_H/{\rm cm^{-2}}\le 22$
is consistent with, or even lower than, all optical survey estimates,
so we do not expect a much \emph{lower} contribution from these sources.

\begin{figure}
\includegraphics[angle=00,scale=1.2]{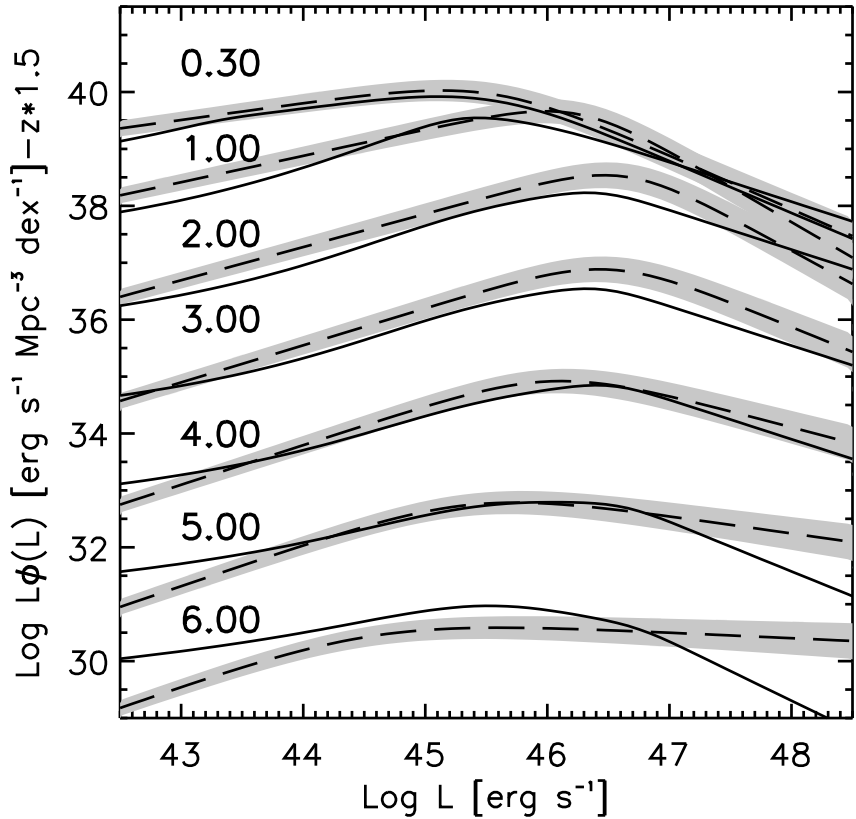}
\caption{Comparison between the bolometric AGN luminosity function
described in \S~\ref{sec|AGNLF} (\emph{solid} line) and the
recently derived estimate by HRH07 (\emph{dashed}
line), shown with its 1-$\sigma$ uncertainty in the bolometric
correction (\emph{gray area}). Curves are vertically offset by $-1.5\times z$.} \label{fig|HopkinsLF}
\end{figure}

HRH07 have recently undertaken an exercise similar to ours,
attempting to construct a bolometric AGN LF that is consistent with
a wide range of data over a large span of energies and wavelengths.
Figure~\ref{fig|HopkinsLF} compares our model to theirs. We find
good agreement in the shape and overall normalization of the AGN LF
in the redshift range $z < 1$ and $3.5 \le z \le 4.5$, while at
intermediate redshifts our model LF is systematically lower. The
normalization difference arises because HRH07 adopt an X-ray
bolometric correction that is about $30\%$ higher (at typical
luminosities) than the Marconi et al. (2004) bolometric correction
used here. If we adopt the HRH07 bolometric correction, then we find
good agreement in the range $1\le z \le 3$, but our model LF is
higher at high and low redshifts. At $z \ge 4$ we adopt a steeper
bright-end LF slope based on the results of GOODs (Fontanot et al.
2006), Cool et al. (2006) and Shankar \& Mathur (2007), who find
evidence for a steeper slope relative to the estimates of Richards
et al. (2006). However, at $1 \le z \le 2.5$ the HRH07 bright-end
slope is steeper than ours, since they effectively require a match
to the SDSS-based measurements of Richards et al. (2004) in this
redshift range, while we retain the shallower slope adopted by U03,
also supported by recent X-ray LF determination of Silverman et al.
(2008). Recent Spitzer studies (Hickox et al. 2007; Polletta et al.
2008) suggest a significant fraction of obscured AGN even at high
luminosities, leaving some room for a slope difference between the
optical and bolometric LFs. We regard the differences between our
model LF and that of HRH07, derived from independent attempts to
match the full range of current data, as a reasonable representation
of the remaining systematic uncertainties in determination of the
bolometric AGN luminosity function. We will show in
\S~\ref{subsec|models} that the difference between these two
determinations does not alter our overall conclusions, but the
differences in normalization and shape at intermediate redshifts do
lead to different preferred accretion parameters for the black hole
population.

\section{The local black hole mass function and the integrated Soltan argument}
\label{sec|LMFSOLTAN}

\subsection{The local black hole mass function}
\label{subsec|LMF}

Marconi et al. (2004) and S04 found similar results for the local
black hole mass function, using somewhat different methods. Both
groups found that starting from the relation between black hole mass
and bulge velocity dispersion ($M_{\bullet}-\sigma$) or the relation
between black hole mass and bulge luminosity ($M_{\bullet}-L_{\rm
sph}$) led to similar local mass functions. However, a more recent
analysis by Lauer et al. (2007a) and Tundo et al. (2007) argues that
these two methods do not provide the same answer. We therefore
revisit some of the uncertainties in computing the local mass
function.

Figure~\ref{fig|LocalMFs} compares the local mass functions obtained
using different relations between black hole mass and host galaxy
properties. Following S04, we start with the galaxy LF in the $r^*$
band by Nakamura et al. (2002), who used light concentration
parameters to distinguish early and late type galaxies and derived
LFs for both. After correcting the Petrosian magnitudes to total magnitudes
by adding $-0.2$ mag, we convert the LF into a LF of spheroids, which we
compute as follows. First we compute the numerical fraction of
Elliptical-S0 (spirals Sab-Sbc-Scd) in the early-type (late-type)
galaxy LF, then correct each morphological galaxy type for its
respective spheroidal luminosity component, as given in Table 1 of
Fukugita et al. (1998) for the $r$-band (see also Yu \& Tremaine
2002 and Marconi et al. 2004). Note that this method is
equivalent to correcting the luminosities themselves adopting an average
weighted fraction of $f_{\rm sph}=0.83-0.85$ for early-type galaxies, and $f_{\rm
sph}=0.27-0.3$ for late-type galaxies, as done in S04 (and references therein).

The solid line in Figure~\ref{fig|LocalMFs} shows our result
using the $M_{\bullet}-L_{\rm sph}$ relation calibrated by McLure \&
Dunlop (2002; their equation 6) for a sample of 18 black holes in
inactive galaxies, where we correct the magnitudes for our
cosmology. Using the calibration in equation (A9) of Tundo et al. (2006)
would give a similar result. These calculations yield
a black hole mass function $\sim$ 34\%
higher than that in S04 (shown with solid squares).
We include a Gaussian scatter of 0.3 dex around the mean
$M_{\bullet}-L_{\rm sph}$ relation in both cases. The intrinsic scatter to insert in the local
relations between black hole mass and host galaxy properties is
empirically uncertain because it is similar in magnitude to the
observational uncertainties themselves. However, most authors
estimate a scatter of about 0.3 dex, which is close to the
value predicted in the numerical simulations of Hopkins et al.
(2007b). In the following we will always adopt this value of the
scatter for all our computations of the local mass function.

If we instead use the galaxy velocity dispersion function and the
$M_{\bullet}-\sigma$ relation in equation (A5) of Tundo et al. (2006),
with a scatter of 0.3 dex, we get the short-dashed line in
Figure~\ref{fig|LocalMFs}, close to the central estimate of S04.
Here we use the velocity dispersion function by Sheth et al. (2003),
which includes their estimate for the contribution of the
bulges in spirals, which in turn accounts for $\sim 25\%$ of the total
estimated black hole mass density. The Sheth et
al. (2003) estimate of the velocity dispersion function is in very
good agreement with the velocity function estimated by S04 through
the bivariate relation of galaxy luminosity and velocity dispersion.
However, using the $M_{\bullet}-\sigma$ relation in Marconi et al.
(2004) yields the dot-dashed line, which is higher than the S04 determination
by about $1-\sigma$. The $M_{\bullet}-\sigma$ relation given in
Ferrarese \& Ford (2005) is steeper than the Tundo et al. (2007)
relation, and adopting it yields a similar mass density to S04, but
a local mass function shifted towards higher masses, shown with a
triple-dot dashed line in Figure~\ref{fig|LocalMFs} (see also Wyithe
2004). Both of these estimates yield a local mass function
below the $M_{\bullet}-L_{\rm sph}$-based local mass function
estimate at $M_{\bullet}<10^{7.5}\, M_{\odot}$, as also noted
by S04.

In the following we will adopt the grey band of Figure~\ref{fig|LocalMFs}, which spans the range of these estimates,
as representative of the mean and the systematic uncertainties of present estimates of the local mass function.
The integrated mass density of the local black hole population is $\rho_{\bullet}=(3.2-5.4)\times 10^5\, {\rm M_{\odot}\, Mpc^{-3}}$ (for $h=0.7$).
Figure~\ref{fig|LocalMFs} also presents three additional estimates of the local mass function. Stars show the results of combining
the Bell et al. (2003) galaxy baryonic mass function (corrected for spheroid fraction as above) with the
H\"{a}ring \& Rix (2004) estimate of the relation between black hole mass and spheroid stellar mass, again assuming
0.3 dex intrinsic scatter. This result is in reasonable agreement with the $M_{\bullet}-\sigma$ estimates, though uncertainties
in the stellar mass-to-light ratio are a remaining source of systematic uncertainty. The dotted curve shows the estimate
of Hopkins et al. (2007b), based on a ``fundamental plane'' relation between black hole mass, galaxy velocity dispersion, and spheroid half-light
radius; it is more sharply peaked than the grey band and implies a higher integrated black hole mass density. Open circles
show Graham et al.'s (2007) estimate based on the correlation between black hole mass and the S\'{e}rsic light-profile index
of the galaxy spheroid. This estimate is similar to Hopkins et al.'s (2007b) near the peak, but it implies an even sharper fall off towards
low black hole masses. Further discussion of the discrepancies in local mass function estimates, and possible routes
to alleviate them, will appear in Shankar \& Ferrarese (in preparation).

\subsection{The integrated mass density}
\label{subsec|massdensity}

Before turning to the evolution of the differential black hole mass
function, the central theme of this paper, we briefly revisit the
classic So{\l}tan (1982) argument, which relates the integrated
black hole density to the integrated emissivity of the AGN
population. If the average efficiency of converting accreted mass
into bolometric luminosity is $\epsilon \equiv L/\dot{M}_{\rm
inflow}c^2$, where $\dot{M}_{\rm inflow}$ is the mass accretion
rate, then the actual accretion onto the central black hole is
$\dot{M}_{\bullet}=(1-\epsilon)\dot{M}_{\rm inflow}$, where the factor $1-\epsilon$ accounts for the fraction of
the incoming mass that is radiated away instead of being added to
the black hole. The rate at which mass is added to the black hole mass function
is then given by
\begin{equation}
\frac{d\rho_{\bullet}}{dt}=\frac{1-\epsilon}{\epsilon
c^2}\int_0^{\infty}\Phi(L)Ld\log L\, .
    \label{eq|soltan}
\end{equation}
The mass growth rate implied by equation~(\ref{eq|soltan}) and our
estimate of the AGN LF from \S 2 is shown by the solid line in
Figure~\ref{fig|SFR-Rhoz}a. We set the radiative efficiency to a
value of $\epsilon=0.075$, as it provides a cumulative mass density
in agreement with the median estimate of the local mass density discussed in
\S~\ref{subsec|LMF}. At each time
step we integrate equation~(\ref{eq|soltan}) down to the observed
faint-end cut in the $2-10$ keV AGN LF, which we parameterize as
\begin{equation}
\log L_{\rm MIN, 2-10\, {\rm keV}}(z)=\log L_{0,\, 2-10\, {\rm
keV}}+2.5\log(1+z)\, .
    \label{eq|Lbolz}
\end{equation}
We set $\log L_{2-10\, \rm{keV}}/({\rm erg\, s^{-1}})=41.5$, in agreement with
the faintest low redshift objects observed by U03 and
La Franca et al. (2005). For a typical $L_{\rm opt}/L_X$, equation~(\ref{eq|Lbolz})
yields an optical luminosity of $M_B\sim -22$ at $z\sim 6$,
comparable to the faintest AGN sources observed by Barger et al.
(2003) in the 2 Msec Chandra Deep Field North (see also
Figure~\ref{fig|BolLF} and Shankar \& Mathur 2007). At each time
step we compute the minimum observed luminosity given in
equation~(\ref{eq|Lbolz}) and convert it into a bolometric quantity
$L_{\rm MIN}(z)$ applying the adopted bolometric correction by Marconi et al. (2004).

Dashed and dot-dashed lines in Figure~\ref{fig|SFR-Rhoz}a show two
recent estimates of the cosmic star formation rate (SFR) as a
function of redshift, from Hopkins \& Beacom (2006), reported with
its 3-$\sigma$ uncertainty region (dark area), and Fardal et al.
(2007). We have multiplied both estimates by a redshift-independent factor of
$8\times 10^{-4}$. Since local
estimates imply a typical ratio $M_{\bullet}/M_{\rm star}\sim
1.6\times 10^{-3}$ for spheroids (e.g., Har\`{\i}ng \& Rix 2004), this is a
reasonable scaling factor if roughly $50\%$ of star formation goes
into spheroidal components (see also Marconi et al. 2004 and Merloni
et al. 2004). The agreement between the inferred histories of black
hole growth and star formation suggests that the two processes are
intimately linked. In particular this association seems to hold down to the last several Gyrs,
even at $z\lesssim 1.5$ when
disk galaxies are expected to dominate the SFR. A possible link
between black hole growth and star formation in disks could arise from re-activations
induced by tidal interactions between satellite
and central galaxies (e.g., Vittorini et al. 2005). Also, bars could possibly funnel
gas into the central black hole, though empirical studies cast
some doubt on this mechanism as a primary trigger for black hole growth (Peeples
\& Martini 2006 and references therein).

Figure~\ref{fig|SFR-Rhoz}b presents the same comparison in
integrated form (see also De Zotti et al. 2006 and Hopkins et al.
2006b). Solid squares show the black hole mass density
obtained by converting the $z=1$ and $z=2$ galaxy stellar mass
function into a black hole mass function by assuming a ratio
$M_{\bullet}/M_{\rm star}$ equal to the local one (i.e., $1.6\times 10^{-3}$). The galaxy
stellar mass function has been computed from the Caputi et al.
(2006) $K$-band galaxy luminosity function, assuming an average
mass-to-light ratio $M_{\rm star}/L_K$=$0.4$ at $z=1$ and
$M_{\rm star}/L_K$=$0.3$ at $z=2$. The latter values have been
obtained from the \emph{Pegase2} code (Fioc \& Rocca-Volmerange
1997) by taking a short burst of star formation ($< 10^9$ yr) and a
Kennicutt double power-law stellar Initial Mass Function. The quoted
values for $M_{\rm star}/L_K$ can be taken as lower
limits, as other choices of the parameters in the code would tend to
increase their value. However, we also note that our result on the
stellar mass function is in good agreement with the recent
estimate by Fontana et al. (2006). Our scaling factor of $1.6\times 10^{-3}$ implicitly assumes
that all the stellar mass in the luminous galaxies probed by these high-redshift observations resides in
spheroidal components today, and is therefore associated with black hole mass.

Figure~\ref{fig|SFR-Rhoz} suggests that the ratio of black hole
growth to SFR is approximately the same at all redshifts, and
suggests a close link between black hole accretion and star
formation. If the average ratio of black hole mass to stellar mass
were much higher than the local value at $z=1-2$, then the squares
in Figure~\ref{fig|SFR-Rhoz}b would shift above the black hole mass
density curve $\rho_{\bullet}(z)$. Increasing $M_{\rm star}/L_K$
would move the squares higher still, while associating only a
fraction of the high-redshift stellar mass with present day
spheroids would move them lower. Our conclusions are in marginal
agreement with those of McLure et al. (2006) and Shields et al.
(2006) and in some disagreement with those of Peng et al. (2006),
who find $M_{\bullet}/M_{\rm star}$ in gravitationally lensed quasar
hosts at $z\sim 1-2$ a factor $\sim 3$ above the local value.
However, several observational biases may effect studies of the
$M_{\bullet}/M_{\rm star}$ ratio in high redshift quasar samples
(Lauer et al. 2007b). Uncertainties in the local normalization of
$\rho_{\bullet}$ and in the stellar mass-to-light ratios still leave
a fair amount of wiggle room, but our results in
Figure~\ref{fig|SFR-Rhoz} disfavor models in which black holes
``grow first'' and spheroid star formation ``catches up'' at later
times. Alternatively, if the average efficiency $\epsilon$ of black
hole accretion is lower at high redshifts, then the
$\rho_{\bullet}(z)$ curve in Figure~\ref{fig|SFR-Rhoz}b would shift
upwards at these redshifts accordingly.

\section{Evolving the black hole mass function}
\label{sec|BHevol}

\subsection{Method}
\label{subsec|method}

Our goal is to calculate the evolution of the black hole mass
function implied by the bolometric AGN LF described in \S 2, given
assumptions about the radiative efficiency and the typical accretion
rate. In the following we will use the symbol $\Phi(x)$ to denote mass and
luminosity functions in logarithmic units of $L$ or $M_{\bullet}$,
i.e.
\begin{equation}
\Phi(x)=n(x)x \ln(10)\, ,
    \label{eq|Phi_x}
\end{equation}
where $n(x)$ is the comoving space density of black holes in the
mass or luminosity range $x\rightarrow x+dx$, in units of ${\rm
Mpc^{-3}}$ for $h=0.7$.

We define the Eddington accretion rate to be
\begin{equation}
\dot{M}_{\rm Edd}\equiv \frac{L_{\rm Edd}}{0.1\, c^2}\simeq
22\left(\frac{M_{\bullet}}{10^9\, M_{\odot}}\right)\, M_{\odot}\,
{\rm yr^{-1}}
    \label{eq|Medd}
\end{equation}
and the dimensionless accretion rate
$\dot{m}=\dot{M_{\bullet}}/\dot{M}_{\rm Edd}$, where $L_{\rm Edd}$
is the standard Eddington luminosity (for Thomson scattering opacity and pure hydrogen
composition) at mass $M_{\bullet}$. The black hole growth rate $\dot{M}_{\bullet}$ is related
to the large scale accretion rate by
$\dot{M}_{\bullet}=(1-\epsilon)\dot{M}_{\rm inflow}$, where
$L=\epsilon \dot{M}_{\rm inflow}c^2$, because a fraction $\epsilon$
of the mass is radiated away before entering the black hole. We define
$f=\epsilon/(1-\epsilon)$, and $f_{0.1}=f/0.1$, so that a black hole
of mass $M_{\bullet}$ growing at a dimensionless rate $\dot{m}$ has
bolometric luminosity
\begin{equation}
L=\epsilon \dot{M}_{\rm inflow}c^2=0.1 f_{0.1}
\dot{M}_{\bullet}c^2=f_{0.1}\dot{m}lM_{\bullet}\, ,
    \label{eq|Lbol}
\end{equation}
where $l\equiv L_{\rm Edd}/M_{\bullet}=1.26\times 10^{38}\, {\rm
erg\, s^{-1}\, M_{\odot}^{-1}}$. Note that we define $\dot{M}_{\rm
Edd}$ for $f=0.1$, so the \emph{mass} Eddington ratio $\dot{m}$ is
linked to the \emph{luminosity} Eddington ratio $\lambda$ by the
radiative efficiency, i.e.
\begin{equation}
\lambda\equiv \frac{L}{L_{\rm Edd}}=\dot{m}f_{0.1}\, .
    \label{eq|Lambda}
\end{equation}
A black hole
accreting at a constant value of $\dot{m}$ grows exponentially in
time with a timescale $t_{\rm
growth}=M_{\bullet}/\dot{M}_{\bullet}=t_s/\dot{m}=4.5\times 10^7\,
\dot{m}^{-1}\, {\rm yr}$, where $t_s$ is equal to the Salpeter
(1964) timescale for $f_{0.1}=1$.

The evolution of the black hole mass function $n(M_{\bullet},t)$ is
governed by a continuity equation (Cavaliere et al. 1973,
Small \& Blandford 1992)
\begin{equation}
\frac{\partial n}{\partial t}(M_{\bullet},t)=-\frac{\partial
(M_{\bullet}\langle \dot{m}\rangle n(M_{\bullet},t))}{\partial
M_{\bullet}}\,,
    \label{eq|conteq}
\end{equation}
where $\langle \dot{m}\rangle$ is the mean dimensionless accretion
rate (averaged over the active and inactive populations) of the black holes of mass $M_{\bullet}$ at time $t$. This
evolution is equivalent to the case in which every black hole grows constantly at the mean accretion rate $\langle \dot{m}\rangle$. In
practice, individual black holes turn on and off, and there may be a
dispersion in $\dot{m}$ values, but the mass function evolution
depends only on the mean accretion rate as a function of mass.

All models in this paper assume a single accretion rate
$\dot{m}=\dot{m}_0$. At any given time, a black hole is either
accreting at $\dot{m}_0$ or not accreting. In some models we allow the
characteristic accretion rate $\dot{m}_0$ to depend on $z$, or on $M_{\bullet}$. The assumption of a
single $\dot{m}$ is clearly not valid for low luminosity AGNs in the
nearby universe, which have a wide range of Eddington ratios (e.g., Heckman et al. 2004, Greene \& Ho 2007).
However, Kollmeier et al. (2005) find that luminous AGN at
$0.5<z<3.5$ have a narrow range of Eddington ratios, with a peak at
$\lambda\sim 1/4$ and a dispersion of 0.3 dex. Since this dispersion
includes contributions from random errors in black hole mass
estimates and bolometric corrections, the true dispersion should be
even smaller. Netzer et al. (2007) find a similar result, with a slightly
larger dispersion, from a sample centered at $z\sim 2.5$. We will consider models with multiple $\dot{m}$ values
in future work, but single-$\dot{m}$ models (also adopted by, e.g.,
Marconi et al. 2004 and S04) are a good starting point for understanding black hole
growth.

If there is a single accretion rate $\dot{m}_0$, then the duty cycle of black hole
activity (i.e., the probability that a black hole of mass
$M_{\bullet}$ is active at a particular time) is given by the ratio
of luminosity and mass functions,
\begin{equation}
P_0(M_{\bullet},z)=\frac{\Phi(L,z)\left|\frac{d\log L}{d\log
M_{\bullet}}\right|}{\Phi(M_{\bullet},z)}\le 1\, ,
    \label{eq|P0general}
\end{equation}
where $M_{\bullet}=L/(f_{0.1}\dot{m}_0 l)$ is the black hole mass
that corresponds to luminosity $L$. Models with constant
$\dot{m}_0$ and $\epsilon$ have $L\propto
M_{\bullet}$, making the Jacobian factor unity. A physically consistent
model must have $P_0(M_{\bullet},z)\le 1$; there must be enough
black holes to produce the observed luminous AGNs.

Our strategy is to start with an assumed black hole mass function
$n(M_{\bullet},z_i)$ at an initial redshift $z_i$, then track the
characteristic curves $M_{\bullet}(M_{\bullet, i},z)$ of
equation~(\ref{eq|conteq}) by direct integration
\begin{eqnarray}
M_{\bullet}(M_{\bullet, i},z)=\int_{z_i}^{z}\langle
\dot{m}_{\bullet}\rangle
M_{\bullet}(z')\frac{dt}{dz'}dz' \nonumber\\
=\int_{z_i}^{z}\dot{m}_0P_0(M_{\bullet},z')M_{\bullet}(z')\frac{dt}{dz'}dz'
\, .
    \label{eq|mdotav}
\end{eqnarray}
Here $P_0(M_{\bullet},z')$ is given by equation~(\ref{eq|P0general})
with the observed luminosity function, and the evolved
$n(M_{\bullet},z)$ is given by
\begin{equation}
n(M_{\bullet},z)=n(M_{\bullet,i},z_i)\frac{dM_{\bullet,i}}{dM_{\bullet}}\,
,
    \label{eq|nmJac}
\end{equation}
where $M_{\bullet}$ is the black hole mass that corresponds to
initial mass $M_{\bullet,i}$. To put this calculation in more physical terms: we take advantage of the fact
that equation~(\ref{eq|conteq}) is equivalent to having every black
hole grow at the rate $\langle \dot{m}(M_{\bullet},z)\rangle$.
Starting with a set of logarithmically spaced initial values of
$M_{\bullet,i}$, we integrate the masses forward in time with a
mid-point scheme from redshift step $z_j$ to step
$z_{j+1}=z_j-\Delta z$
\begin{equation}
M_{\bullet,j+1}=M_{\bullet,j}+\dot{m}_0P_0(M_{\bullet,j+1/2},z_{j+1/2})M_{\bullet,j+1/2}\frac{dt}{dz}\Delta
z\, ,
    \label{eq|leapfrog}
\end{equation}
where the values at the mid-point $j+1/2$ are evaluated by
extrapolating black hole masses for a half step $\Delta z/2$ using
$dM_{\bullet}/dt$ at the beginning of the step and the duty cycle $P_0$ is computed
using the mid-step mass function and the luminosity function evaluated at
$z_{j+1/2}=z_j+\Delta z/2$.
We choose sufficiently small redshift steps $\Delta z$ such that the
total mass density added at each iteration matches the one obtained
by integration of the bolometric luminosity function
(equation~\ref{eq|soltan}), and we have confirmed that smaller $\Delta z$
yields essentially identical results. Since the number of black
holes is conserved, $n(M_{\bullet},z)\Delta
M_{\bullet}$=$n(M_{\bullet,i},z)\Delta M_{\bullet,i}$.

As an aside, we note that this approach is mathematically
equivalent to solving the continuity equation in the form
\begin{equation}
\frac{\partial n(M_{\bullet},t)}{\partial t}=-\frac{\dot{m}_0}{t_s
\ln(10)^2 M_{\bullet}}\frac{\partial \Phi(L,z)}{\partial \log L}\,
    \label{eq|POsingle}
\end{equation}
used by, e.g. Small \& Blandford (1992) and Marconi et al. (2004).
Equation~(\ref{eq|POsingle}) follows from equation~(\ref{eq|conteq})
with the assumption of a single $\dot{m}_0$ and $f_{0.1}$, and it is
also valid in the case of a redshift dependent $\dot{m}_0$. From
equation~(\ref{eq|POsingle}), it is explicitly clear that the
evolution of the black hole mass function depends only on the
initial conditions, the observed luminosity function, and the values of $\dot{m}_0$
and $f_{0.1}=L/(M_{\bullet}l \dot{m}_0)$, given
the assumptions made here. The So{\l}tan argument
(\S~\ref{subsec|massdensity}) relates the integrated black hole
mass density to the integrated emissivity of the AGN population. The
continuity equation yields the full evolution of the black hole mass
function in terms of the full luminosity function; in essence, it
applies the So{\l}tan argument as a function of mass. We have checked
that our method yields consistent results against direct integration
of the equation.

For our integrations, we generally start at $z_i=6$ and determine
$n(M_{\bullet,i},z_i)$ from equation~(\ref{eq|leapfrog}) with our
estimate of the luminosity function at $z=6$ and an assumed
$P_0=0.5$. This relatively high initial duty cycle implies we
start with nearly the minimal black hole mass function required to
reproduce the luminosity function. By $z\sim 3.5$ the
integration has essentially forgotten the initial value of $P_0$
unless we set it much lower (e.g., $P_0<0.01$), since the accreted mass is much larger
than the mass in the initial black hole population. The clustering of luminous
AGN at $z\sim 3-4$ implies duty cycles of at least several percent (Shen et al. 2007;
Shankar et al., in preparation). In our standard calculations, we set
$P_0$ to zero for masses $M_{\bullet, {\rm MIN}}(z)$ that would
produce AGN luminosities below the minimum observed luminosity at
that redshift (indicated by the cutoffs in Figure~\ref{fig|BolLF});
we do not extrapolate the luminosity function below these minimum
values. Since $L_{\rm MIN}(z)$ drops
towards lower $z$, we must have a supply of black holes in place to
provide AGNs of lower luminosity as these become visible. To ensure
this, we set our initial black hole mass function to
\begin{equation}
n(M_{\bullet,i},z_i)=8\times n(M_{\bullet,{\rm
MIN}}(z_i),z_i)\times\left(\frac{M_{\bullet}}{M_{\bullet,{\rm
MIN}}(z_i)}\right)^{-2.3}\, ,
\label{eq|InitialMF}
\end{equation}
i.e., below $M_{\bullet,\rm MIN}(z_i)$ we boost $n(M_{\bullet},z_i)$
by a factor of $8$ and add an $M_{\bullet}^{-2.3}$ rise (see
Figure~\ref{fig|bestFitModel}, below). This high initial abundance
of low mass black holes ensures that we are never forced to duty
cycles $P_0>1$. Once $M_{\bullet}$ significantly exceeds
$M_{\bullet,{\rm MIN}}(z)$, the black hole mass function is
dominated by growth rather than initial values, and
$n(M_{\bullet},z)$ is insensitive to the assumed
$n(M_{\bullet,i},z_i)$. We adopted this procedure so that our results would not depend
on the extrapolation of the luminosity function into regions where it is not observed.
In practice, we find that extrapolating the luminosity function as a power law down to very
faint luminosities yields similar results. We follow the $L_{\rm min}$ procedure for our standard models in \S\S~\ref{subsec|bestfit} and ~\ref{subsec|models}, and
use extrapolation of the LF in \S\S~\ref{subsec|etamdot}-\ref{subsec|merging} where it gives
more stable numerical solutions.

\subsection{The reference model}
\label{subsec|bestfit}

Figure~\ref{fig|bestFitModel} shows the results of our calculations
for a reference model with our standard estimate of the AGN bolometric
luminosity function and accretion parameters $\dot{m}=0.60$ and $\epsilon=0.065$
($f_{0.1}\sim 0.7$), which yield good overall agreement with the
average estimate of the local black hole mass function.
Panel~\ref{fig|bestFitModel}a shows the evolution of the mass
function starting from the initial condition at $z=6$ (solid
triangles) and continuing through to $z=0$. Because of the
luminosity-dependent density evolution in the observed luminosity
function, the massive end of the black hole mass function builds up
early, and the lower mass regime grows at later redshifts. For
$M_{\bullet}>10^{8.5}\, {\rm M_{\odot}}$, the mass function is
almost fully in place by $z=1$. Panel~\ref{fig|bestFitModel}b plots
$M_{\bullet}\Phi(M_{\bullet})$, proportional to the fraction of
black hole mass per logarithmic interval of $M_{\bullet}$, which
allows better visual comparison to the observed local mass function
and highlights the contribution of each black hole mass bin to
the total mass density at each time. The accretion model agrees well
with our estimate of the local mass function (grey band) up to $\sim 10^{9.3}\, {\rm
M_{\odot}}$. At higher masses our model exceeds the observational
estimate, but in this regime the estimate relies on extrapolation of
the scaling relations, and it is sensitive to the assumed intrinsic scatter.
In addition, the high-mass end of the predicted mass function is sensitive to the
bright end of the LF at $z\lesssim 2$ (see
\S~\ref{subsec|models}).
The open circles with error bars
show the estimate of the local mass function by Graham et al. (2007), which
we cannot reproduce even approximately with constant $\dot{m}_0$.
If this estimate is correct, then the low end of the
luminosity function must be produced mainly by high mass black holes accreting at low $\dot{m}$, so that
the predicted growth of low mass black holes is reduced (see \S~\ref{subsec|dmdtMbh}).

Figure~\ref{fig|bestFitModel}c plots the duty cycle as a function of mass for different
redshifts, as labeled. The duty cycle for $M_{\bullet}\sim 10^9\, {\rm M_{\odot}}$
is $\sim 0.2$ at $z\sim 4-5$, falling to 0.03-0.08 at $z=2-3$, when quasar activity
is at its peak, then dropping to 0.003 at $z=1$ and $\sim 10^{-4}$ at $z=0$. Below $z\sim 3$, the duty cycle
rises towards low black hole masses. This ``downsizing'' evolution, in which high mass black holes
complete their growth early but low mass black holes continue to grow at late times, is required
by the observed LF evolution in any model with approximately constant $\dot{m}_0$. Above $10^9\, {\rm M_{\odot}}$ the duty
cycle curves are generally flat, because the constant slope of the bright end of the LF drives the growth
of a mass function with a parallel slope. Below $z=1$, the LF slope changes, and the duty cycle
curves are no longer flat. By running several cases with varying initial conditions, we find that the duty cycles
for $z\lesssim 3.5$ are insensitive to the assumed initial duty cycle
(which determines the initial mass function) provided $0.01\lesssim P_0(z=6)\le 1$.

Several lines of independent observational evidence have set
constraints on the duty cycle of AGNs at different redshifts.
Porciani et al. (2004), Croom et al. (2005), Porciani and Norberg
(2006), and da Angela et al. (2006) have analyzed large samples of
optical AGNs in the Two Degree Field (2dF) QSO Redshift Survey,
showing that their large-scale ($\gtrsim 1$ Mpc) clustering
properties are consistent with quasars residing in very massive
halos with $M_{\rm h}\sim 5\times 10^{12}-10^{13}\, {\rm
M_{\odot}}$. Comparing the AGN abundance to the abundance of such halos
implies duty cycles of order of $\sim (1-4)\%$ in the redshift range $z\sim 1-2$.
Similar results have been confirmed by large-scale
optical AGN clustering studies in the Sloan Digital Sky Survey
(SDSS; e.g., Myers et al. 2007). Shen et al. (2007) have extended the SDSS
results to $z=3-4$, obtaining similar results for halo masses and duty cycles.
Similar or lower values for the duty cycle could be found directly by
comparing the AGN number density to the galaxy luminosity function
at comparable redshifts, the latter now well estimated and complete
down to rather low luminosities (e.g., Pannella et al. 2006). Using \emph{Chandra} X-ray observations of the Hubble Field
North, Nandra et al. (2002) and Steidel et al.
(2002) find that about 3\% of a large spectroscopic survey of
$\sim 1000$ Lyman Break Galaxies at the average redshift of $z\sim
3$ host an X-ray bright, moderately obscured AGN. Within the uncertainties of the
comparison, these estimates of AGN duty cycles are in good agreement with the predictions
in Figure~\ref{fig|bestFitModel}c. We will present predictions for AGN clustering bias in
\S~\ref{sec|hosts} below, and we will carry out quantitative comparisons to observed clustering in future work.
The occurrence of AGN in massive star-forming galaxies
is much higher than these overall duty cycle estimates
(Alexander et al. 2003; Borys et al. 2005), supporting the scenario in
which star-formation in massive galaxies is closely related to black
hole growth (e.g., Granato et al. 2006).

Figure~\ref{fig|bestFitModel}d shows the black hole accretion
histories as a function of relic mass and redshift, following
equation~(\ref{eq|mdotav}). Tracing back one of these curves shows the mass that a
``typical'' black hole of present mass $M_{\bullet}$ had at earlier
times, as predicted by our model. The dot-dashed line shows the minimum
black hole mass associated with $L_{\rm MIN}(z)$ at each redshift.
By construction, all growth curves are flat (dashed lines) before crossing
$L_{\rm MIN}(z)$. Figure~\ref{fig|bestFitModel}d shows again that high mass black holes build their
mass early and experience little growth at late times, while lower mass black holes grow rapidly at lower
redshifts.


%
%

%
\subsection{Parameter variations}
\label{subsec|models}

The top panels of Figure~\ref{fig|Models} show the effect of varying the two
main model parameters, the
radiative efficiency $\epsilon$ and accretion rate $\dot{m}_0$. Figure~\ref{fig|Models}a
compares models with varying $\epsilon$ and accretion rate fixed to
the reference model value $\dot{m}=0.60$. Lowering (raising) the radiative
efficiency has the effect of shifting the accreted mass function up
(down) and shifting the peak of the mass distribution to higher
(lower) masses. Recall that the luminosity Eddington ratio is $\lambda=\dot{m}f_{0.1}$; if we
held $\lambda$ fixed instead of $\dot{m}$, then curves in Figure~\ref{fig|Models}a would be
vertically displaced with no horizontal shift.
Figure~\ref{fig|Models}b shows models with varying $\dot{m}_0$,
all with $\epsilon=0.065$. The So{\l}tan argument implies that $\rho_{\bullet}=\int M_{\bullet} \Phi(M_{\bullet})d\log M_{\bullet}$
should be the same for models with the same $\epsilon$ that reproduce the same observed LF. The area
under the curves in Figure~\ref{fig|Models}b is thus the same for all models, but the peak in the mass
function shifts to lower $M_{\bullet}$ for higher $\dot{m}_0$ because a given observed luminosity
is associated with a lower black hole mass. While
there are still systematic uncertainties in the normalization and
shape of the local mass function, the peak of the
$M_{\bullet}\Phi(M_{\bullet})$ distribution (Figure~\ref{fig|LocalMFs})
is in good agreement among all the
estimates, and this, in turn, sets strong constraints on
$\dot{m}_0$. Shifting the peak of the accreted mass distribution
to $\log (M_{\bullet}/M_{\odot})\sim 8$ would require
Super-Eddington accretion rates, while shifting it to $\log (M_{\bullet}/M_{\odot})\sim 9$
requires $\dot{m}_0\sim 0.2-0.3$.
Changes to $\epsilon$ and $\dot{m}_0$ shift the accreted
mass function in different directions in the $M_{\bullet}\Phi(M_{\bullet})-M_{\bullet}$
plane, but they do not alter its shape.


The lower panels of Figure~\ref{fig|Models} show the effect of changing the input LF. In panel~\ref{fig|Models}c,
solid and dotted curves show models with no Compton-thick sources or a double fraction of Compton-thick sources, respectively,
with radiative efficiency and accretion rate fixed at the reference model values $\epsilon=0.065$ and $\dot{m}_0=0.6$.
The predicted black hole mass functions are similar in shape but offset in amplitude by about $\pm$ 20\%. These
offsets are similar to our estimated uncertainty in the local mass function, shown by the grey band. We
can restore agreement between the double Compton-thick model and the central estimate of the local mass function
by raising $\epsilon$ to 0.08 (to produce more luminosity with the same black hole mass) and lowering
$\dot{m}_0$ to 0.5 (to compensate the influence of higher $\epsilon$ on the peak location). Conversely,
the no-Compton-thick model yields similar predictions if $\epsilon$ is lowered to 0.05 and $\dot{m}_0$
is raised to 0.8. As noted in \S~\ref{sec|AGNLF}, we find better agreement with the observed
X-ray background for no Compton-thick and worse agreement for double Compton-thick (see Figure~\ref{fig|XRBGs}).

Figure~\ref{fig|Models}d compares our reference model to the
one obtained by integrating the AGN bolometric LF of HRH07.
For the reference model values $\epsilon=0.065$ and $\dot{m}_0=0.6$,
the predicted mass function shifts to higher normalization and higher peak mass (solid line).
Principally this difference reflects the higher bolometric correction adopted by HRH07,
which shifts their LF to higher normalization at $1\lesssim z \lesssim 3$ (Figure~\ref{fig|HopkinsLF}).
The HRH07 estimate also has higher peak luminosity in this redshift range, leading to higher peak mass.
Adopting a higher efficiency and accretion rate, $\epsilon=0.09$ and $\dot{m}_0=1$, largely
compensates these two effects. As noted in \S~\ref{sec|AGNLF}, the HRH07 LF has a steeper bright-end slope
at intermediate redshifts, and this leads to a steeper high-mass slope of the accreted mass function, improving
agreement with observational estimates above $M_{\bullet}\sim 10^9\, {\rm M_{\odot}}$.

To present our model comparison in a more global way, we characterize the local mass function
by four quantities that contain complementary information:
\begin{itemize}
  \item The integrated black hole mass density $\rho_{\bullet}=\int M_{\bullet}\Phi(M_{\bullet})d\log M_{\bullet}$.
  \item The location $\log M_{\rm PEAK}$ at which the mass function peaks in the $M_{\bullet}\Phi(M_{\bullet})-M_{\bullet}$ plane.
  \item The width of the mass function, characterized by $(\Delta \log M_{\bullet})_{1/2}$ such that twice
  the integral from $\log M_{\rm PEAK}$ to $\log M_{\rm PEAK}+(\Delta \log M_{\bullet})_{1/2}$ contains half the total
  mass density, $2\times \int_{\log M_{\rm PEAK}}^{\log
\bar{M_{\bullet}}} \Phi(M_{\bullet})M_{\bullet}d\log
  M_{\bullet}$$=1/2\times \rho_{\bullet}$. If $M_{\bullet}\Phi(M_{\bullet})$ were a Gaussian, $(\Delta \log M_{\bullet})_{1/2}$
  would be $\sim 0.26$ times the full width at half maximum.
  \item The asymmetry of the mass function, characterized by $\Delta
  \rho_{\bullet}/\rho_{\bullet}$, the normalized difference in the mass density integrated above and below $\log M_{\rm PEAK}$.
\end{itemize}

Figure~\ref{fig|multiparam} plots the predicted values of these four quantities
as a function of the radiative efficiency $\epsilon$ for models with our standard
AGN LF and accretion rates $\dot{m}_0=1, 0.6, 0.45, 0.3$. Grey horizontal bands
show the observational estimates corresponding to the grey band in Figure~\ref{fig|LocalMFs}.
Short-dashed and long-dashed horizontal lines show instead the parameters for the local mass functions inferred by
Graham et al. (2007) and Hopkins et al. (2007b), respectively (see Figure~\ref{fig|LocalMFs}).
As expected from the So{\l}tan argument, the predicted $\rho_{\bullet}$ is proportional
to $(1-\epsilon)/\epsilon$ but independent of $\dot{m}_0$. Consistency with the
grey band observational estimate requires $0.063\le \epsilon \le 0.1$.\footnote{Our reference model
of \S~\ref{subsec|bestfit} has $\epsilon$ at the low end of this range, rather
than the middle, because we chose parameters to match the mass function near its peak, where it is
best determined. However, the reference model predicts $\rho_{\bullet}=5.24\times 10^5\, {\rm M_{\odot}\, Mpc^{-3}}$,
above the middle of the grey band.} Reproducing the
observed value of $\log M_{\rm PEAK}/{\rm M_{\odot}}\approx 8.5$ then requires $0.4\lesssim \dot{m}_0 \lesssim 0.8$,
though for any given $\epsilon$ the value of $\dot{m}_0$ is determined to $\sim 10\%$.
As noted in our discussion of Figure~\ref{fig|Models}a, the location of the mass function
peak is determined by $\lambda=L/L_{\rm Edd}$ largely (but not completely) independent of $\epsilon$,
and reproducing the observed $\log M_{\rm PEAK}$ implies $\lambda\approx 0.4-0.5$ over the range
$0.05\lesssim \epsilon \lesssim 0.13$.

The higher $\rho_{\bullet}$ implied by the Hopkins et al. (2007b) local mass function requires a lower efficiency,
$\epsilon\approx 0.06$ which in turn implies a higher $\dot{m}_0\approx 1$ to match $\log M_{\rm PEAK}$. (Note that this is
\emph{not} the comparison in Figure~\ref{fig|Models}d, where we use the HRH07 \emph{luminosity} function
but show our standard estimate of the local \emph{mass} function). Our inferred ranges for $\epsilon$ and
$\dot{m}_0$ are consistent with the findings of Marconi et al. (2004) and S04, who obtained ($\epsilon$, $\lambda$)=(0.08, 0.5)
and (0.09, 0.3), respectively.
Note that S04 tends towards slightly higher values of the radiative efficiency mainly because they
used an X-ray bolometric correction normalized to the optical bolometric correction of Elvis et al. (1994), i.e., $C_B$=11.8,
while we have adopted the lower value suggested by recent work (see \S~\ref{sec|AGNLF}).

With $\epsilon$ and $\dot{m}_0$ fixed by matching $\rho_{\bullet}$ and $\log M_{\rm PEAK}$, there is no freedom
in the model to adjust the predicted width and asymmetry of the mass function, but these prove remarkably
insensitive to $\epsilon$ and $\dot{m}_0$ in any case. To substantially alter these predictions one must
either change the input luminosity function or add a new physical ingredient to the model, such as
mass- or redshift-dependent $\dot{m}_0$, mergers, or multiple accretion rates. For our minimal
model and standard LF estimate, the predicted widths are larger than any of the observational
estimates. This discrepancy is largely a consequence of the shallow high-mass slope of the predicted mass function
(see Figure~\ref{fig|Models}), which in turn is sensitive to the bright-end slope of the luminosity function
in the range $1\lesssim z \lesssim 3$, as discussed earlier. Note, however, that the observational estimates of
$(\Delta \log M_{\bullet})_{1/2}$ depend on the extrapolation of the black hole-galaxy correlations above $M_{\bullet}\approx 10^9\, {\rm M_{\odot}}$,
where they are quite uncertain, so the grey band in Figure~\ref{fig|multiparam} should not necessarily
be taken as a true upper limit. The predicted asymmetries are in reasonable agreement with the observational estimates,
except for the Graham et al. (2007) mass function, in which the population of low mass black holes is greatly reduced.

Figure~\ref{fig|multiparamLF} presents a similar comparison for models with different input luminosity functions.
Here we fix $\dot{m}_0=0.75$ in all cases, noting that the value of $\dot{m}_0$ affects the predicted
$\log M_{\rm PEAK}$ but has little impact on other quantities. Eliminating Compton-thick sources from
our standard LF (dotted curve) lowers the efficiency required to match a given $\rho_{\bullet}$ by about
$20\%$, to $\epsilon=0.062$ for $\rho_{\bullet}=4.26\times 10^5\, {\rm M_{\odot}\, Mpc^{-3}}$ (short-dashed line).
Doubling the Compton-thick contribution (not shown) has a similar effect in the opposite direction,
raising the required $\epsilon$ by 20\%. Changing the Compton-thick contribution has little
impact on $\log M_{\rm PEAK}$ for a given $\epsilon$, or on the width or asymmetry of the predicted mass function.
Eliminating \emph{all} obscured sources and retaining only Type I quasars drastically reduces the predicted $\rho_{\bullet}$,
so such a model (which is implausible on direct observational grounds anyway) can only be reconciled
with the local black hole population if the efficiency is very low or if the true
value of $\rho_{\bullet}$ is significantly below our observational estimate. However,
a model in which the full bolometric luminosity function is just three times that of Type I AGN --- two obscured
sources for every unobscured source independent of luminosity and redshift --- yields similar predictions to our full model
with the luminosity-dependent column density distribution of U03.

For the HRH07 luminosity function, we require
$0.079\lesssim \epsilon \lesssim 0.125$ to reproduce the grey-band range of $\rho_{\bullet}$, $25\%$ higher than our standard
LF at fixed $\rho_{\bullet}$. As discussed earlier, this difference mainly reflects the higher bolometric correction
adopted by HRH07, which boosts the normalization of the LF at intermediate redshifts. The HRH07 LF also leads to a higher
$M_{\rm PEAK}$ at a given $\epsilon$, and matching $\log M_{\rm PEAK}/{M_{\odot}}\approx 8.5$ implies
$0.7\lesssim \dot{m}_0 \lesssim 1.1$ for the range of $\epsilon$ that matches $\rho_{\bullet}$. The HRH07
LF predicts a somewhat narrower and more asymmetric mass function, mainly because of the steeper bright-end
slope at intermediate redshifts, which steepens the high mass end of the black hole mass function. These results
accord with those shown already in Figure~\ref{fig|Models}d. If we adopt \emph{both} the HRH07 luminosity function
and the Hopkins et al. (2007b) black hole mass function, which has $\rho_{\bullet}=5.7\times 10^5\, {\rm M_{\odot}\, Mpc^{-3}}$
and $\log M_{\rm PEAK}/{\rm M_{\odot}}\approx 8.4$, we find $\epsilon=0.075$ and $\dot{m}_0=1.3$.

The remaining uncertainties in the luminosity function and the local black hole mass function still leave considerable
uncertainty in the radiative efficiency. If we assume that our no-Compton-thick LF and the HRH07 LF bracket
the luminosity function uncertainty and our grey band brackets the $\rho_{\bullet}$ uncertainty, the allowed range
is $0.05\lesssim \epsilon \lesssim 0.125$, and there is clearly some room for the luminosity function or $\rho_{\bullet}$
to be outside this range. However, the high efficiencies predicted
by MHD simulations of thick accretion disks around rapidly spinning black holes, $\epsilon\approx 0.15-0.2$
(e.g., Gammie et al. 2004), can only represent the mean efficiency of black hole accretion if $\rho_{\bullet}$
is considerably below our estimates or the luminosity function is substantially higher than even the HRH07 estimate. There are some
$\rho_{\bullet}$ estimates as low as $2-3\times 10^5\, {\rm M_{\odot}\, Mpc^{-3}}$ (e.g., Wyithe et al. 2004). Our conclusions about efficiency are in agreement with most
previous studies, though Elvis et al. (2002) reached a different conclusion mainly because they assumed that the sources producing
the X-ray background have an effective mean redshift of $z\sim 2$, while subsequent data have shown that the peak contribution to the background
is at $z\lesssim 1$ (e.g., U03; La Franca et al. 2005).

\subsection{Redshift-dependent Eddington ratio}
\label{subsec|etamdot}

So far we have assumed that all black holes at all times radiate at
a constant, mildly sub-Eddington rate, an assumption supported by
Kollmeier et al. (2005)'s results for luminous quasars in the AGES survey at $0.5\lesssim z \lesssim 4$.
However,  Eddington ratios of local AGN have a wide distribution, with a large fraction of Seyferts radiating
at $\lambda < 0.1$. Some studies also show evidence for
mass-dependence of the typical Eddington ratio (e.g., Heckmann et al. 2004; Vestergaard 2004; McLure \& Dunlop 2004; Constantin \& Vogeley 2006;
Dasyra et al. 2007; Shen et al.\ 2007b), though systematic uncertainties in the
reverberation mapping techniques and extrapolations of empirical
virial relations (e.g., Kaspi et al. 2000; Bentz et al. 2006)
make it hard to draw firm conclusions.
In this section and the one that follows, we
discuss simple models with $\dot{m}_0$ values (and thus Eddington ratios)
that depend on redshift or mass. More realistic models would incorporate a \emph{range}
of Eddington ratios with the relative importance of high and low accretion rates depending on mass or
redshift; we will investigate such models in future work. We note again that the integrated
black hole mass density should depend only on $\epsilon$ (equation~\ref{eq|soltan}), but
the shape of the black hole mass function can change if $\dot{m}_0$ is not constant.

Figure~\ref{fig|etamdot} compares the results of our reference model, which has $\epsilon=0.065$,
$\dot{m}_0=0.6$, to a model with the same $\epsilon$ but
\begin{equation}
\dot{m}_0(z)=[1-\exp(-z/z_s)]\, ,
    \label{eq|mdotz}
\end{equation}
with $z_s=2$. This implies values of $\dot{m}_0=0.78,0.63,0.39,0.22,0.095, 0.049$ at
$z=3,2,1,0.5,0.2,0.1$, respectively, which are consistent with the
average drop shown in Figure 6b of Vestergaard et al. (2004) for
the more massive black holes, assuming the distribution starts at
$\dot{m}_0\approx 1$ at very high redshifts. The decreasing $\dot{m}_0(z)$ significantly
alters the shape of the black hole mass function (Figure~\ref{fig|etamdot}a) by shifting the low
redshift growth of the mass density away from low mass black holes and towards higher mass black
holes. This model still requires a substantial number of low-mass black holes residing in spiral galaxy
bulges if it is to match the local mass function.

Figures~\ref{fig|etamdot}b compares the duty cycles produced by the
$\dot{m}_0(z)$ model to those from the reference model, at several redshifts.
The reference model duty cycles are
strongly decreasing functions of redshift and black hole mass, dropping by up
to two orders of magnitude in the lower redshift bins. The
$\dot{m}_0(z)$-model ``damps'' the downsizing, leaving a much milder
variation of the duty cycle over time and mass. As emphasized by SW03, the downward shift of the LF break luminosity at $z<2$ can be explained either
by a strong suppression of activity in high mass black holes or by a drop in characteristic
Eddington ratios. Our $\dot{m}_0(z)$ model is a compromise between these two explanations;
higher mass black holes still build their mass earlier than low mass black holes, but the difference
is smaller than in the reference model (Figure~\ref{fig|etamdot}c).

Greene \& Ho (2007) have
recently estimated the local MF for active black holes through the
virial relations mentioned above. When compared with the local mass function of
relic black holes, their study supports a
duty-cycle of $P_0\sim 10^{-2.35}$ for black holes with mass $\log M_{\bullet}/M_{\odot}\gtrsim 7$,
consistent with the $z\sim 0$ duty-cycle inferred from this model,
and about two orders of magnitude above the one corresponding to
our reference model (Figure~\ref{fig|etamdot}b).

\subsection{Eddington ratio varying with black hole mass}
\label{subsec|dmdtMbh}

In the previous section we studied a model in which the Eddington ratio steadily decreases
with redshift and is fixed with black hole mass. Here we
consider the complementary model in which the Eddington ratio distribution is
constant with redshift but declines with black hole
mass.

We adopt
\begin{equation}
\dot{m}_0(M_{\bullet})=0.445\times \left(\frac{M_{\bullet}}{10^9\,
{\rm M_{\odot}}}\right)^{0.5}\,
    \label{eq|EddMbh}
\end{equation}
with a constant radiative efficiency $\epsilon=0.075$.
As discussed in \S~\ref{subsec|method},
we evolve this model by integrating equation~(\ref{eq|POsingle}) with a fixed black hole mass grid, extrapolating
the luminosity function below $L_{\rm MIN}(z)$. Relative to the reference model, the $\dot{m}_0(M_{\bullet})$ model associates
the peak of the LF with higher mass black holes, especially at lower redshifts. The low redshift growth
of low mass black holes is strongly suppressed, and the $z=0$ abundance of low mass black holes is much lower,
consistent with the Graham et al. (2007) local mass function estimate, as shown in Figure~\ref{fig|dmdtMbh}.
The dependence of duty cycle on mass is also much stronger in this model because
of the paucity of low mass black holes; note that the duty cycle at low masses
is high but the amount of growth is low because of the low accretion rates. Getting our mass-dependent
$\dot{m}_0$ model to run at all requires a high initial black hole density, with a
duty cycle $P_0(z=6)=0.01-0.02$; otherwise, the slow growth of low mass black holes leads to required
duty cycles that exceed unity at intermediate redshifts. Alternatively, we can start with a higher initial
duty cycle and only ``turn on'' the mass dependence of equation~(\ref{eq|EddMbh}) after an epoch
of growth at constant $\dot{m}_0$.

More generally, we can vary the Eddington ratio both in redshift and mass,
$\dot{m}_0(M_{\bullet},z)=T(z)B(M_{\bullet})$.
The continuity equation can then be written
\begin{eqnarray}
\frac{\partial n(M_{\bullet},t)}{\partial t}=-\frac{T(z)}{(\ln 10)^2M_{\bullet}}\times\,\,\,\,\,\,\,\,\,\,\,\,\,\,\,\,\,\,\,\,\,\,\,\,\,\,\,\,\,\,\\
\nonumber \frac{\partial}{\partial \log M_{\bullet}}
\nonumber\left[B(M_{\bullet})\Phi(L,z)\left|\frac{d \log
\nonumber L}{d\log M_{\bullet}}\right|\right]=
\nonumber -\frac{T(z)}{\ln(10)^2M_{\bullet}}\times\,\,\,\,\,\,\,\,\,\,\,\,\,\,\,\,\,\,\,\,\,\,\\
\nonumber  \left[B(M_{\bullet})\frac{\partial
\nonumber \Phi(L,z)}{\partial \log L}\left(\frac{\partial\log L}{\partial\log
\nonumber M_{\bullet}}\right)^2+\left|\frac{\partial \log L}{\partial \log M_{\bullet}}\right|
\nonumber \Phi(L,z)\frac{\partial B(M_{\bullet})}{\partial \log M_{\bullet}}\right] \, ,
\label{eq|MeddMbh}
\end{eqnarray}
where we have set to zero the \emph{second} derivative $\partial^2 \log L/(\partial \log M_{\bullet})^2$
assuming a power-law relation like that in equation~(\ref{eq|EddMbh}).
We have used equation~(19)
to investigate models in which we vary the Eddington ratio with redshift and mass
as given in equations~(\ref{eq|mdotz}) and ~(\ref{eq|EddMbh}).
We find that we can get somewhat better fits to Graham et al.'s (2007) local
mass function, since both the implemented trends
act to decrease the numbers of low mass black holes.


\subsection{Black hole mergers}
\label{subsec|merging}

Observations show, and hierarchical galaxy formation models predict, that a significant
fraction of galaxies experience mergers with comparably massive galaxies during their
lifetime. At least some of these galaxy mergers are likely to be accompanied by mergers of
the central black holes that they contain, though the mechanisms that shrink
black hole orbits to the scale where gravitational radiation can drive a final
merger are not fully understood (see Merritt \& Milosavljevi\'{c} 2005).
In a future paper, we will
will combine our accretion model with theoretically
predicted merger rates for cold dark matter subhalos (Yoo et al. 2007)
to create models with realistic contributions of accretion and merger driven growth.
Here we illustrate the potential impact
of mergers on the black hole mass function using a simple mathematical model
that assumes constant probability
of equal mass mergers per Hubble time,
similar to the models of Richstone et al. (1998) and SW03.
We will ignore physically interesting complications such as ejection of black holes
by gravitational radiation or three-body interactions or variations of radiative efficiency
caused by the impact of mergers on the black hole
spin distribution (e.g., Hughes \& Blandford 2003; Volonteri et
al. 2003; Gammie et al. 2004; Islam et al. 2004; Yoo \& Miralda-Escud\'{e} 2004;
Merritt \& Milosavljevi\'{c} 2005; Volonteri et
al. 2006). These effects, along with unequal mergers and the mass and redshift-dependence
of merger rates, should be considered in a complete model of the evolving
black hole population. Mergers redistribute mass within the black hole population,
but they do not change the integrated mass density, so they do not affect
the integrated So{\l}tan (1982) argument. (In principle, gravitational radiation
during mergers can \emph{reduce} the integrated mass density [see Yu \& Tremaine 2002], but we do
not consider this effect here).

%

In our model, a black hole of mass $M_{\bullet}$ has a probability
$P_{\rm merg}$ of merging with an equal mass black hole in the
Hubble time $t_H(z)$ (age of the Universe at redshift $z$).
Therefore the fraction $F$ of black holes that merge in a timestep
$\Delta t$ is
\begin{equation}
F=P_{\rm merg}\times\frac{\Delta t}{t_H(z)}\, .
    \label{eq|mergF}
\end{equation}
At each time $t_1$, we first advance the mass function to time
$t_2=t_1+\Delta t$ with accretion only, then add to each bin of the mass function
an increment (which may be positive or negative)
\begin{equation}
\Delta \Phi(M_{\bullet},t_2)=\frac{F\times \Phi
\left(\frac{M_{\bullet}}{2},t_2\right)}{2}-F\times
\Phi(M_{\bullet},t_2)\, ,
    \label{eq|merging}
\end{equation}
where the second term represents black holes lost from the bin by merging
and $\Phi \left(\frac{M_{\bullet}}{2},t_2\right)$ is calculated by
interpolation.

Figure~\ref{fig|merging} shows the evolution of the black hole mass function for a model
with $P_{\rm merg}=0.5$ and our reference model values of
$\epsilon=0.065$ and $\dot{m}_0=0.6$. Comparing the $z=0.02$
output to that of the reference model shows that the net effect of
merging is to slightly lower the abundance of black holes below the peak
of the mass function and to significantly increase the abundance of very massive
black holes, as expected
(see also Malbon et al. 2006; Yoo et al. 2007).
Although we start merging at $z=6$, comparison of Figure~\ref{fig|merging}
to Figure~\ref{fig|bestFitModel}b shows that the mass function is nearly identical
to that of the reference model down to $z=1$. Mergers at this level
only change the mass function noticeably once
accretion-driven evolution has slowed.

Since our reference model already produces an excess of massive
black holes relative to local estimates of the black hole function,
adding mergers only makes the match to observations worse. However,
the impact of mergers is evident mainly at $M_{\bullet}>10^9\, {\rm
M_{\odot}}$, where the local mass function estimates are most
sensitive to the adopted scatter in the black hole-host scaling
relations and to the extrapolations of these relations to the most
luminous galaxies. Accommodating the predictions of our merger model
would require an intrinsic scatter of $\sim 0.5$ dex at high masses,
or a change in slope of the scaling relations. Either of these
changes seems possible given the current observational
uncertainties, but neither seems likely (see, e.g., discussions by
Wyithe 2004; Batcheldor et al. 2006; Lauer et al. 2006; Tundo et al.
2007). If we adopt the HRH07 LF in place of our standard LF, then
the predicted black hole mass function \emph{with} mergers is
similar to that of our reference model \emph{without} mergers (see
Figure~\ref{fig|merging}). The impact of mergers at this level is
therefore comparable to the systematic uncertainties associated with
the AGN luminosity function.

In fact, the value $P_{\rm merg}=0.5$ is high compared to
theoretical predictions for massive galaxies (Maller et al. 2006),
and theoretically predicted merger rates decline towards lower
masses. Observational estimates of the galaxy merger rate and its
mass dependence span a substantial range (see, e.g., Bell et al.
2006; Conroy et al. 2007; Masjedi et al. 2008, and references
therein), and $P_{\rm merg}=0.5$ is roughly consistent with the high
end of these estimates. We conclude that the impact of mergers on
the black hole mass function is probably small compared to remaining
uncertainties in accretion-driven growth, except perhaps for the
rare, high mass black holes. The most interesting impact of black
hole mergers may arise indirectly, through their effect on black
hole spins and thus on radiative efficiencies (e.g., Volonteri et
al. 2005).
%

\subsection{Tabulation of luminosity and mass functions}
\label{subsec|Tables}

For the convenience of readers who may wish to use them, we provide
electronic tables that list our estimate of the AGN bolometric
LF and some of our model predictions of the black hole mass function and duty cycle, all
as a function of redshift. Table~\ref{Table|FitsLF} lists
our standard LF estimate, the LF after eliminating or doubling
the number of Compton thick sources, and the HRH07 LF estimate evaluated at the same
redshifts and luminosities for convenient comparison.
Table~\ref{Table|Fits} lists the predicted black hole mass function at the same
redshifts for our reference model values of $\epsilon=0.065$, $\dot{m}_0=0.60$,
computed for each of the luminosity functions in Table~\ref{Table|FitsLF}.
The final column lists the mass function predicted for the HRH07 LF and the parameter
values $\epsilon=0.09$ and $\dot{m}_0=1$, which yield a good match to the local mass function
for this LF. Table~\ref{Table|DutyCycles} lists instead the duty cycles corresponding to the same models
reported in Table~\ref{Table|Fits}, computed at the same redshifts and black hole masses.
[Prior to publication, the full tables can be found in electronic format at
{\tt{http://www.astronomy.ohio-state.edu/\\
$\sim$shankar/Models}}.]

\section{Space density and bias of AGN hosts}
\label{sec|hosts}

Predictions of the black hole mass function at redshifts $z>0$ can be tested observationally
if one assumes that black hole mass increases monotonically with the luminosity
of the host galaxy or the mass of its parent halo. This assumption is unlikely
to be perfect, but given the tight correlation between black hole mass
and bulge mass observed today, it may be a reasonable approximation. Figure~\ref{fig|hosts}a shows,
at several redshifts, the predictions of our reference model for the cumulative
space density of galaxies that can host an AGN of luminosity $L$ or greater. Since
the model assumes a single $\dot{m}_0$, this is simply equal to the cumulative
space density of black holes of mass $M_{\bullet}>L/(f_{0.1}\dot{m}_0l)=1.9\times 10^8\, {\rm M_{\odot}}
\times (L/{\rm 10^{46}\, {\rm erg\, s^{-1}}})$. If the monotonic assumption holds, then the observed
space density of galaxies brighter than $L_{\rm host}(L_{\rm AGN})$ should equal
the space density predicted in Figure~\ref{fig|hosts}a, where $L_{\rm host}(L_{\rm AGN})$
is the luminosity of galaxies that host AGN of luminosity $L_{\rm AGN}$.

As emphasized by Haiman \& Hui (2001) and Martini \& Weinberg (2001), the clustering
of AGN can be a powerful diagnostic for the duty cycle of black hole activity: a
high duty cycle implies that halos of AGN are rare, hence high mass, hence strongly clustered.
In the context of our models, we can predict the halo mass $M_h$ associated with black holes
of mass $M_{\bullet}$ by matching the cumulative mass functions,
\begin{equation}
\Phi(>M_{\bullet},z)=\Phi(>M_h,z)\, .
    \label{eq|nhosts}
\end{equation}
Here $\Phi(>M_{\bullet},z)$ is the space density of black holes more massive than $M_{\bullet}$
at redshift $z$, in units of ${\rm Mpc^{-3}}$, as predicted by our evolutionary model,
and $\Phi(>M_h,z)$ is the space density of halos more massive than $M_h$ expected for a $\Lambda$CDM
cosmological model, which we compute using the Sheth \& Tormen (1999) halo mass function with
cosmological parameters $\Omega_m=0.3, \Omega_{\Lambda}=0.7, h=0.7, \sigma_8=0.8$ and linear power spectrum
taken from Smith et al. (2003) with $\Gamma=0.2$. This determination assumes both a monotonic
relation between black hole mass and halo mass and one black hole per halo, but we have checked
that allowing black holes to reside in subhalos does not substantially change our results if
we adopt the subhalo statistics of Vale \& Ostriker (2004).
Finally, we compute the bias of black holes of mass $M_{\bullet}$, and thus the bias of
AGN of luminosity $L=f_{0.1}\dot{m}_0 l M_{\bullet}$, using
Sheth et al.'s (2001) analytic model for halo bias,
\begin{equation}
b(L,z)=b(M_{\bullet},z)=b(M_h,z)\, .
    \label{eq|bias}
\end{equation}
Figure~\ref{fig|hosts}b shows the predicted bias as a function of luminosity
$L$ at several redshifts for the reference model of \S~\ref{subsec|bestfit}.
The predicted bias is much stronger at high redshift because the comoving space
density of black holes is lower (Figure~\ref{fig|bestFitModel}) and because
the bias of halos of a given space density is higher. Points in Figure~\ref{fig|hosts}b
show recent observational estimates of AGN clustering bias from the 2dF
and SDSS quasar redshift surveys (da Angela et al. 2007; Myers et al. 2007; Porciani \& Norberg 2007)
in the redshift range $0.8\lesssim z \lesssim 1.3$ and $1.7\lesssim z \lesssim 2.1$ (Porciani \& Norberg 2007)
and from SDSS at $3\lesssim z \lesssim 3.5$
(Shen et al. 2007a). Our reference model is quite successful at predicting
the absolute level of bias (and thus AGN clustering strength) for quasars with $L\ge 10^{46}\, {\rm erg\, s^{-1}}$ at $z\sim 1$,
$z\sim 2$ and $z\sim 3$, and, especially, at explaining the strong trend of bias among these samples.
It appears to underpredict the clustering
of lower luminosity AGN at $z\sim 1$, though Myers et al. (2007) note the including
the possible impact of stellar contamination on their sample expands their error bars by a factor
of $\sim 1.5$ (an effect we have not included in Figure~\ref{fig|hosts}b).

We reserve a detailed examination of AGN clustering constraints --- including an assessment of what models
can be ruled out with present data or improved measurements --- for future work. Allowing scatter
between black hole mass and halo mass or a range of Eddington ratios can significantly
alter predictions both for AGN host density and for AGN clustering,
so these classes of models will be especially interesting to explore.
For example, the stronger clustering in the Myers et al. (2007)
sample relative to our reference model predictions
could indicate that a fraction of these lower luminosity AGN
are associated with high mass black holes radiating at low Eddington ratios
(e.g., Lidz et al. 2006).

\section{Discussion}
\label{sec|DiscuConclu}

We have constructed self-consistent models for the evolution of the supermassive
black hole population and the AGN population, in which black holes
grow at the rate implied by the observed luminosity function given assumed values
of the radiative efficiency $\epsilon$ and the characteristic
accretion rate $\dot{m}_0=\dot{M}_{\bullet}/\dot{M}_{\rm Edd}$ (see Eqs.~\ref{eq|conteq}, ~\ref{eq|mdotav}, ~\ref{eq|POsingle}).
These models can be tested against the mass distribution of black holes in the local universe, and they make predictions
for the duty cycles of black holes as a function of redshift and mass, which
can be tested against observations of quasar hosts and quasar clustering if one assumes
an approximately monotonic relation between the masses of black holes and the masses of their host galaxies and halos.
Our method is similar to that used previously by Cavaliere et al. (1971), Small \& Blandford (1992),
Yu \& Tremaine (2002), SW03, Marconi et al. (2004), and S04. However, we have drawn on more recent
data on the AGN luminosity function and the local black hole mass function, and we have considered a broader
spectrum of models. This approach can be considered a ``differential'' generalization of the So{\l}tan (1982) argument
relating the integrated emissivity of the quasar population to the integrated mass
density of the local black hole population.

Our model for the bolometric AGN luminosity function starts from the model of U03, based on X-ray
data from a variety of surveys, but we adjust its parameters as a function of redshift in light
of more recent measurements and data at other wavelengths. In agreement with previous
studies we find that the LF of optical AGNs is roughly consistent with that
of X-ray AGN that have absorbing column densities $\log N_H/{\rm cm^{-2}}\le 22$ and that
unobscured AGN dominate the bright end of the LF. We show that the latest constraints on the
hard X-ray background ($E\sim 10-100$ keV) from \emph{INTEGRAL} and from the \emph{PDS} instrument
on BeppoSax support a reduced normalization relative to extrapolations
from other missions at lower energies.
They therefore favor a lower contribution from very highly
obscured AGN ($\log N_H/{\rm cm^{-2}}\gtrsim 24.5$) than some previous estimates
(but see also Gilli et al. 2007). Our estimate of the bolometric
AGN LF is independent in implementation but similar in spirit to that of HRH07.
The most significant differences for purposes of this investigation
are that HRH07 have a higher LF normalization at $z\sim 1-3$, principally because of their choice of bolometric correction,
and they have a steeper bright-end slope at $z\sim 1-2.5$, where they adopt
the slopes measured by Richards et al. (2006) from the SDSS and we use the slopes inferred
by U03 from X-ray data. We regard the differences between our estimate and that of HRH07 as a reasonable
indication of the remaining systematic uncertainties in the bolometric LF of AGNs.

With our LF estimate, the bolometric emissivity of the AGN population tracks recent estimates of the cosmic star
formation rate as a function of redshift.
(Comparisons based on the space density of high luminosity quasars [e.g., Richards et al. 2006,
Osmer 2004 and references therein]
reach a different conclusion because at low redshifts the bright
end of the AGN LF drops much more rapidly with time than the overall emissivity.)
The integrated black hole mass density implied by this emissivity
is $\sim 8\times 10^{-4}$ of the stellar
mass at all redshifts, or about half of the estimated ratio
of black hole mass to bulge stellar mass in local galaxies.
This tracking favors scenarios in which black holes and the stellar mass of bulges grow
in parallel, with about $50\%$ of the star formation linked to black hole growth
at all redshifts. These findings are hard to reconcile with any models where
black hole growth substantially precedes stellar mass buildup, or with recent claims
that the ratio of black hole mass to stellar mass is much larger at high redshift than the local value.
However, our finding refers to integrated densities, so it does not indicate
the relative timing of black hole and bulge growth on an object-by-object basis.

\begin{figure}
\includegraphics[angle=00,scale=1.]{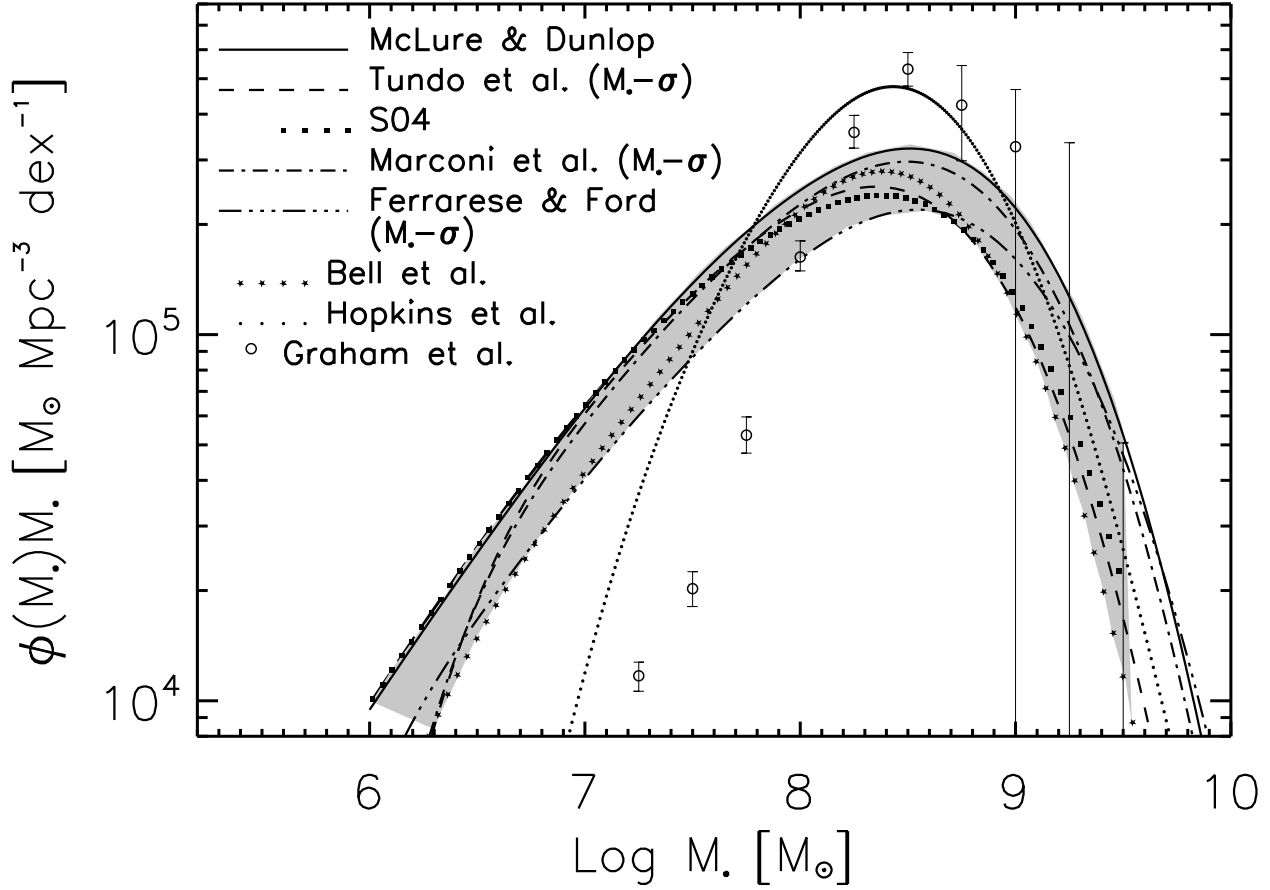}
\caption{Comparison among estimates of the local black hole
mass function. Lines show estimates using various calibrations of the $M_{\bullet}-L_{\rm sph}$,
$M_{\bullet}-\sigma$, or $M_{\bullet}-M_{\rm star}$ relations as described in the text,
assuming a 0.3-dex intrinsic scatter in all cases. The grey band
encompasses the range of these estimates. Filled small circles show the
determination of Hopkins et al. (2007b) using the black hole
``fundamental plane'' and open circles show the determination of Graham et al. (2007)
using the relation between black hole mass and S\'{e}rsic index.} \label{fig|LocalMFs}
\end{figure}

\begin{figure}
\includegraphics[angle=00,scale=1.1]{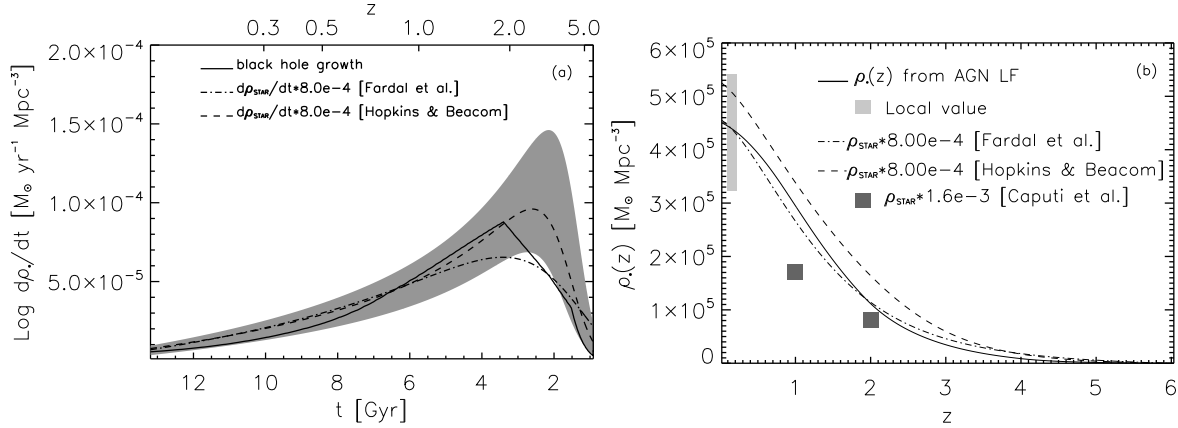}
\caption{Black hole growth and stellar mass growth. (\emph{a}) Average black hole accretion rate as
computed via equation~(\ref{eq|soltan}) compared to the SFR as given
by Hopkins \& Beacom (2006) and Fardal et al. (2007),
scaled by the factor $M_{\bullet}/M_{\rm STAR}=0.5\times
1.6\times 10^{-3}$. The grey area shows the 3-$\sigma$ uncertainty region
from Hopkins \& Beacom (2006). (\emph{b})
Cumulative black hole mass density as a function of redshift. The solid line is the prediction
based on the bolometric AGN luminosity function. The
\emph{light-gray area} is the local value of the black hole mass
density with its systematic uncertainty as given in
Figure~\ref{fig|LocalMFs}. The \emph{dark squares} are estimates of
the black hole mass function at $z=1$ and $z=2$ obtained from the
stellar mass function of Caputi et al. (2006) and Fontana et al.
(2006), scaled by the local ratio $M_{\bullet}/M_{\rm STAR}=1.6\times
10^{-3}$. The lines are the integrated stellar mass densities based on the SFR histories
in panel (\emph{a}), scaled by $8\times 10^{-4}$.} \label{fig|SFR-Rhoz}
\end{figure}

\begin{figure}
\includegraphics[angle=00,scale=1.17]{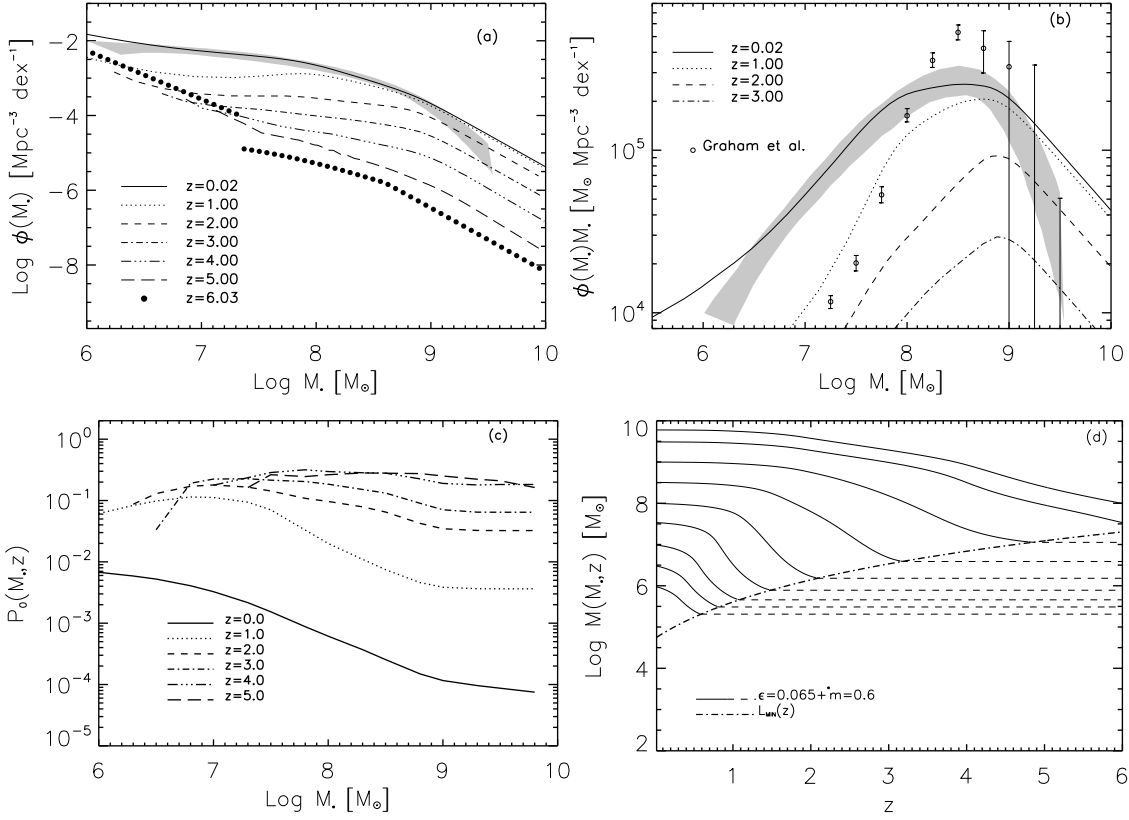}
\caption{Results for a reference model that agrees well with the
local black hole mass function, with $\epsilon=0.065$, $\dot{m}_0=0.60$,
and an initial duty cycle $P_0(z=6)=0.5$. (\emph{a}) The accreted mass function shown at different
redshifts as labeled, compared with the local mass function
(\emph{grey area}). (\emph{b}) Similar to (\emph{a}) but plotting
$M_{\bullet}\Phi(M_{\bullet})$ instead of $\Phi(M_{\bullet})$.
(\emph{c}) Duty-cycle as a function of black hole mass and redshift as labeled.
(\emph{d}) Average accretion histories for black holes of
different relic mass $M_{\bullet}$ at $z=0$ as a function of
redshift; the \emph{dashed} lines represent the curves when the
black hole has a luminosity below $L_{\rm MIN}(z)$ and therefore
does not grow in mass; the minimum black hole mass corresponding to
$L_{\rm MIN}(z)$ at different redshifts is plotted with a
\emph{dot-dashed} line.} \label{fig|bestFitModel}
\end{figure}

\begin{figure}
\includegraphics[angle=00,scale=1.]{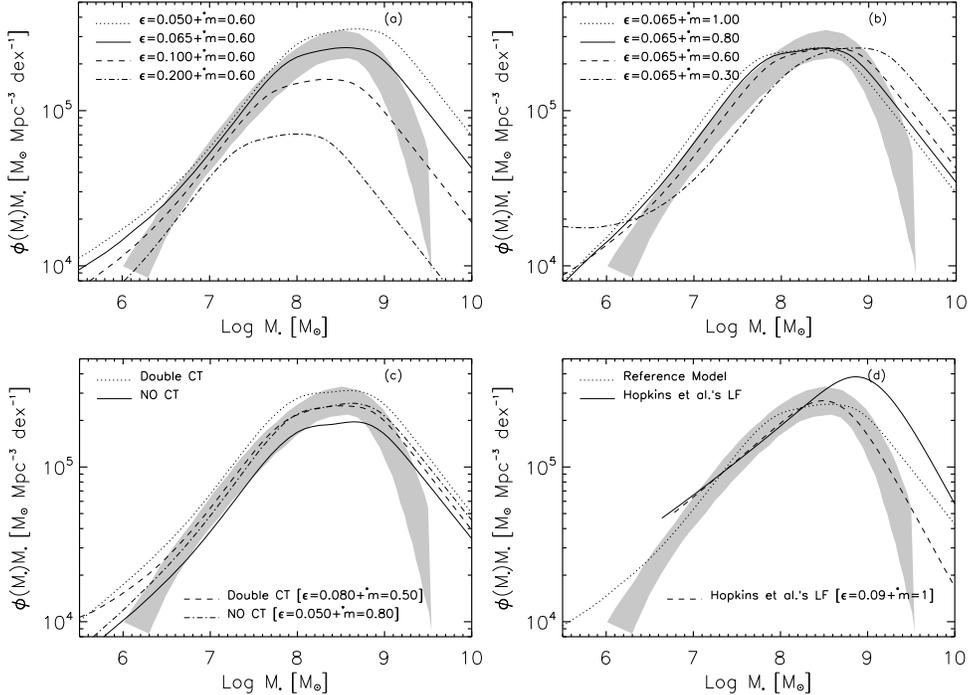}
\caption{Effect of model inputs on the predicted black hole
mass function at $z=0$. (\emph{a}) Varying $\epsilon$ at fixed $\dot{m}_0=0.6$.
(\emph{b}) Varying $\dot{m}_0=$ at fixed $\epsilon=0.065$. (\emph{c})
\emph{Solid} and \emph{dotted} curves show the effect of removing or doubling the fraction of Compton-thick
sources, keeping $\epsilon=0.065$ and $\dot{m}_0=0.6$ fixed at their reference values.
\emph{Dashed} and \emph{dot-dashed} curves show these modified models with parameter values chosen to reproduce
the local mass function. (\emph{d}) The {\it solid}
curve shows the effect of adopting the
HRH07 LF, with $\epsilon=0.065$ and $\dot{m}_0=0.6$. The {\it dashed}
curve shows a model
with the HRH07 LF and parameter values $\epsilon=0.09$ and $\dot{m}_0=1$.} \label{fig|Models}
\end{figure}

\begin{figure}
\includegraphics[angle=00,scale=1.]{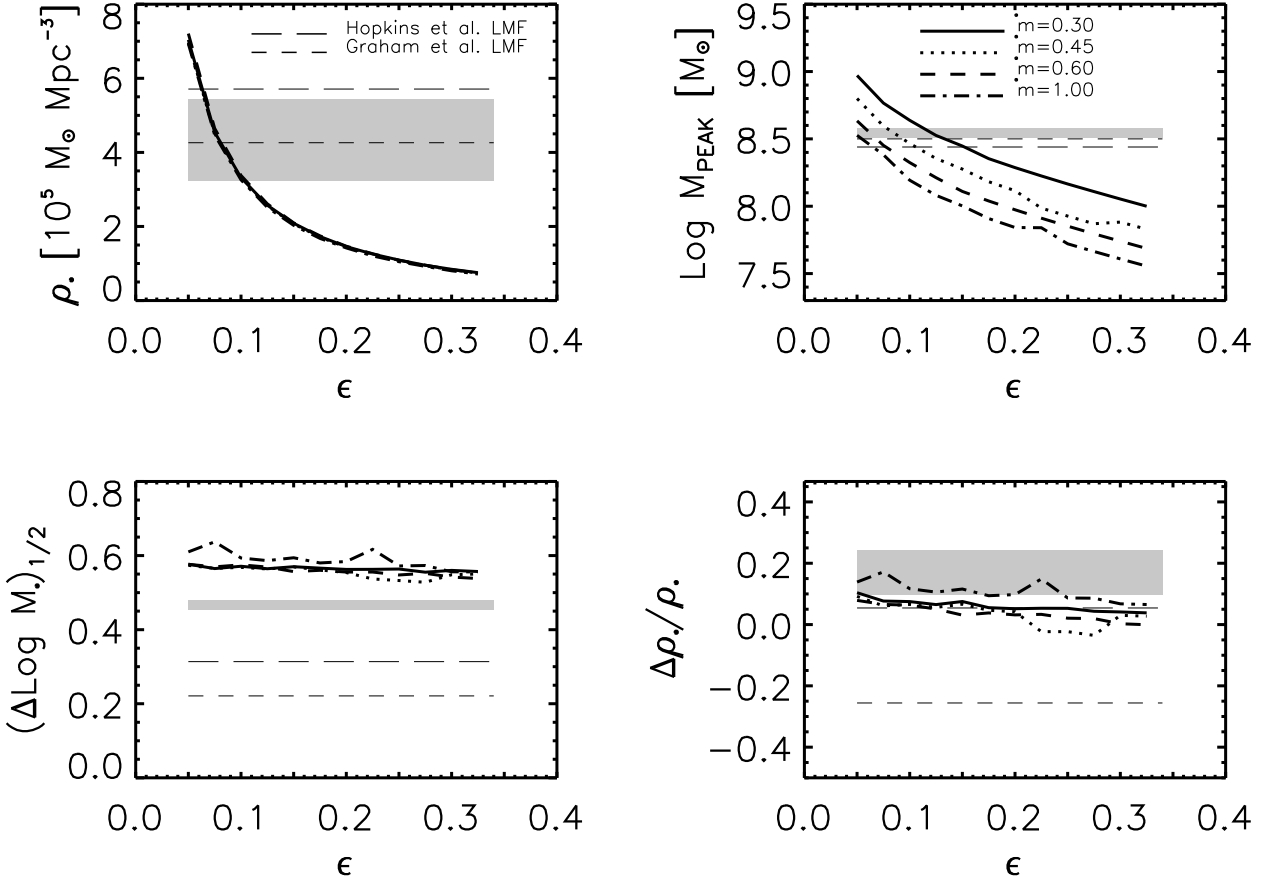}
\caption{Models incorporating our standard luminosity function estimate compared to
the properties of the local black hole mass function described in \S~\ref{subsec|models}: integrated
mass density ($\rho_{\bullet}$, \emph{upper left}), peak mass ($M_{\rm PEAK}$,
\emph{upper right}), width ($M_{\bullet, 1/2}$,
\emph{lower left}), and asymmetry ($\Delta
\rho_{\bullet}/\rho_{\bullet}$, \emph{lower right}). The \emph{thick}
lines show model predictions as a function of radiative efficiency $\epsilon$, for several values
of $\dot{m}_0$ as labeled. The horizontal \emph{grey band} shows observational estimates of the four
quantities based on the grey band in Figure~\ref{fig|LocalMFs}. Horizontal \emph{long-dashed} and \emph{short-dashed}
lines show the values derived from the local mass functions of Hopkins et al. (2007b) and
Graham et al. (2007), respectively.} \label{fig|multiparam}
\end{figure}

\begin{figure}
\includegraphics[angle=00,scale=1.]{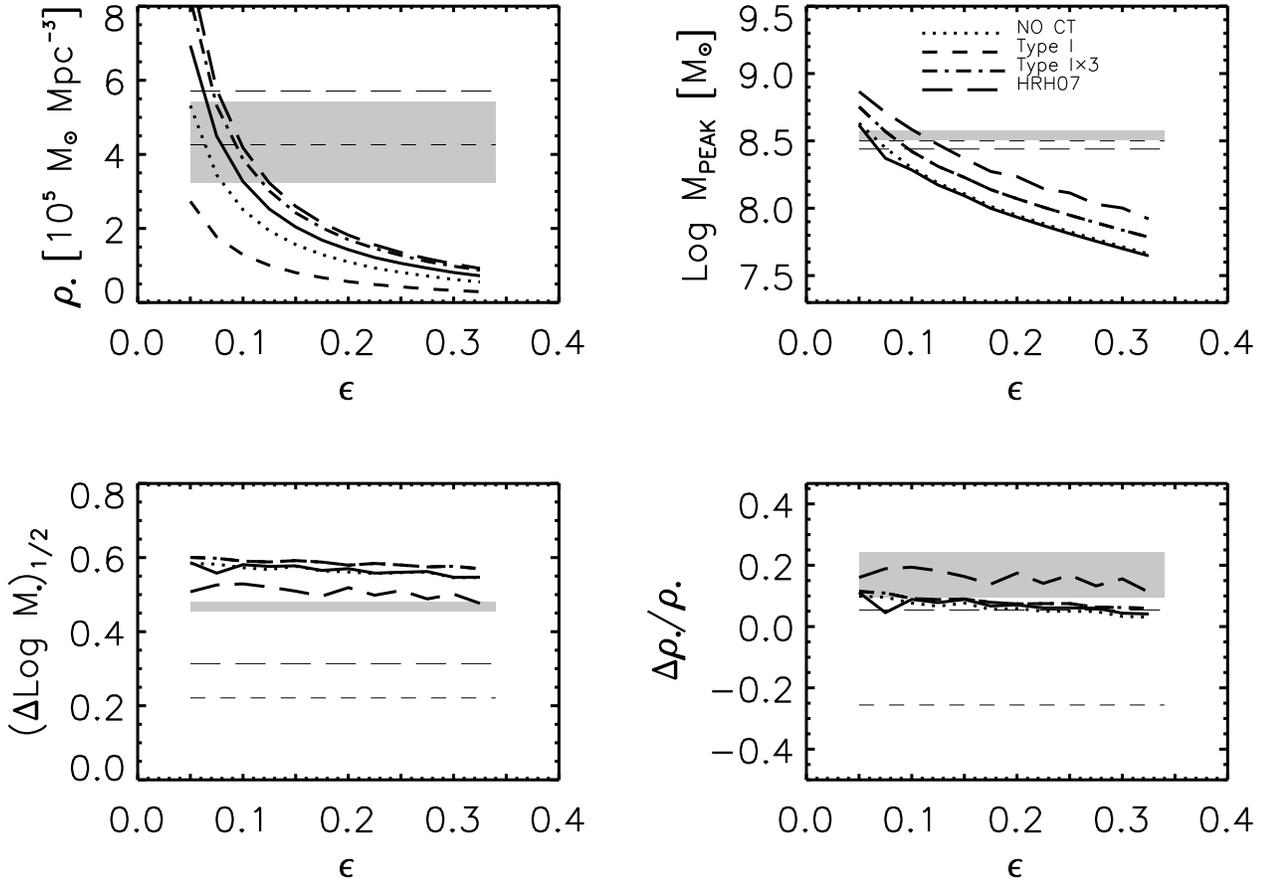}
\caption{Local black hole mass function parameters predicted for different LF inputs.
The accretion rate is
fixed to $\dot{m}_0=0.75$ in all cases. The \emph{thick} lines refer to models
with our standard LF estimate (\emph{solid} line), with no contribution of Compton-thick sources (\emph{dotted} line),
with only Type I AGNs included (\emph{dashed} line), with Type I
AGNs multiplied by a constant factor of 3 at all redshifts and all
luminosities (\emph{dot-dashed} line), and with the HRH07 LF (\emph{long-dashed} line).
Grey band and horizontal thin lines are as
in Figure~\ref{fig|multiparam}.} \label{fig|multiparamLF}
\end{figure}

\begin{figure}
\includegraphics[angle=00,scale=1.17]{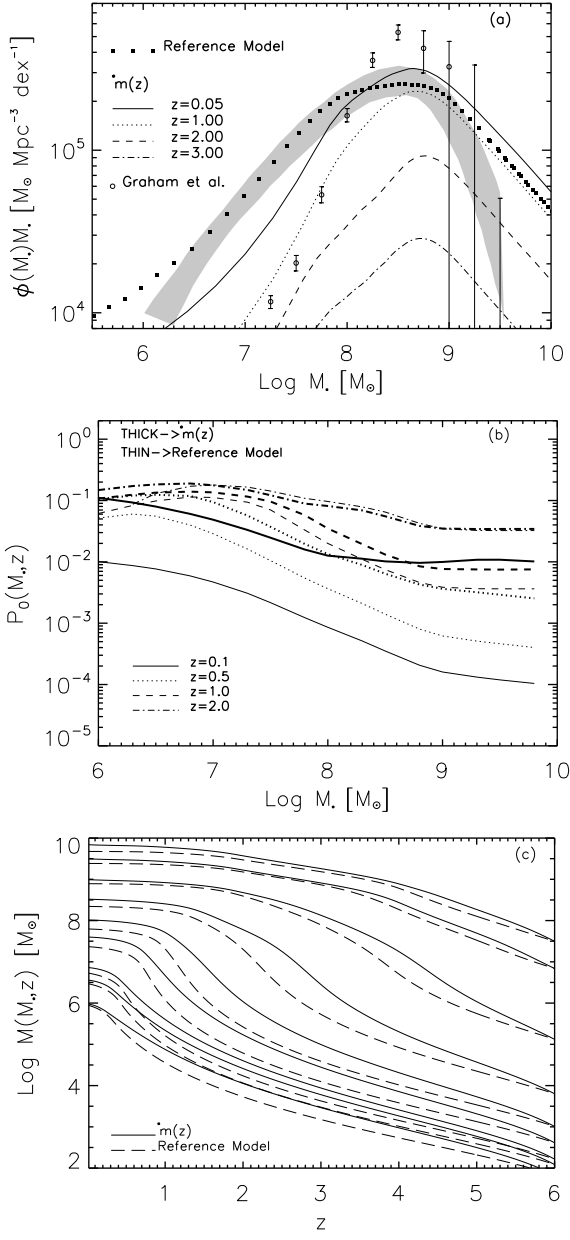}
\caption{Model with $\epsilon=0.065$ and decreasing $\dot{m}_0(z)$ as given in
equation~(\ref{eq|mdotz}), compared to our reference model with constant $\dot{m}_0=0.6$
and $\epsilon=0.065$.
Panels (a), (b), (c) show, respectively, the evolution of the mass function, the evolution
of the duty cycle, and mass growth along characteristics, in the same format as Figure~\ref{fig|bestFitModel}.
Reference model results are shown by \emph{squares} in (a), by
\emph{thinner lines} in (b), and by \emph{long-dashed lines} in (c).}
\label{fig|etamdot}
\end{figure}

\begin{figure}
\includegraphics[angle=00,scale=1.3]{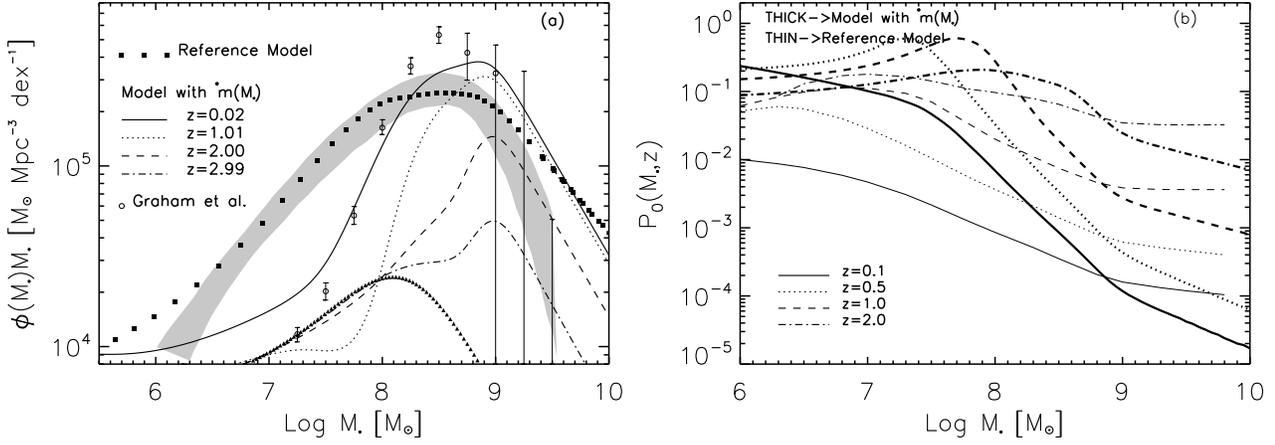}
\caption{Model with $\epsilon=0.075$ and a mass-dependent Eddington
ratio $\dot{m}_0=0.445(M_{\bullet}/10^9\, {\rm M_{\odot}})^{0.5}$
(equation~\ref{eq|EddMbh}), compared to our reference model. The
format is similar to panels (a) and (b) of
Figure~\ref{fig|etamdot}.} \label{fig|dmdtMbh}
\end{figure}

\begin{figure}
\includegraphics[angle=00,scale=1.]{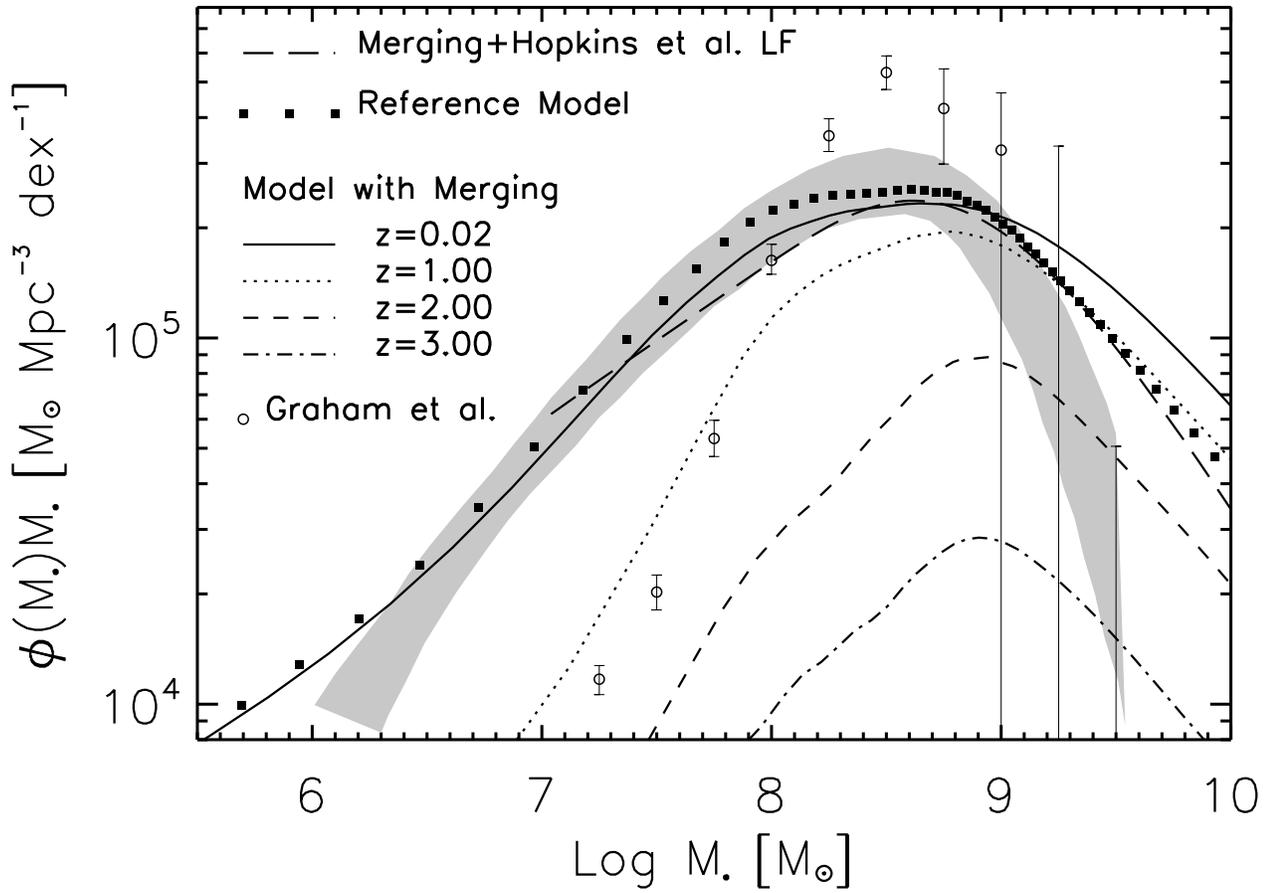}
\caption{Evolution of the black hole mass function in a model with black
hole mergers. Accretion-driven growth is computed assuming $\epsilon=0.065$,
$\dot{m}_0=0.60$, and each black hole has a probability $P_{\rm merg}=0.5$
of merging with another black hole of equal mass per Hubble time $t_H(z)$.
Squares show the $z=0$ predictions of the reference model (same accretion
parameters, no merging), and the long-dashed line shows the $z=0$ mass function
for a merger model with the HRH07 LF and accretion parameters $\epsilon=0.09$,
$\dot{m}_0=1$.} \label{fig|merging}
\end{figure}

\begin{figure}
\includegraphics[angle=00,scale=1.3]{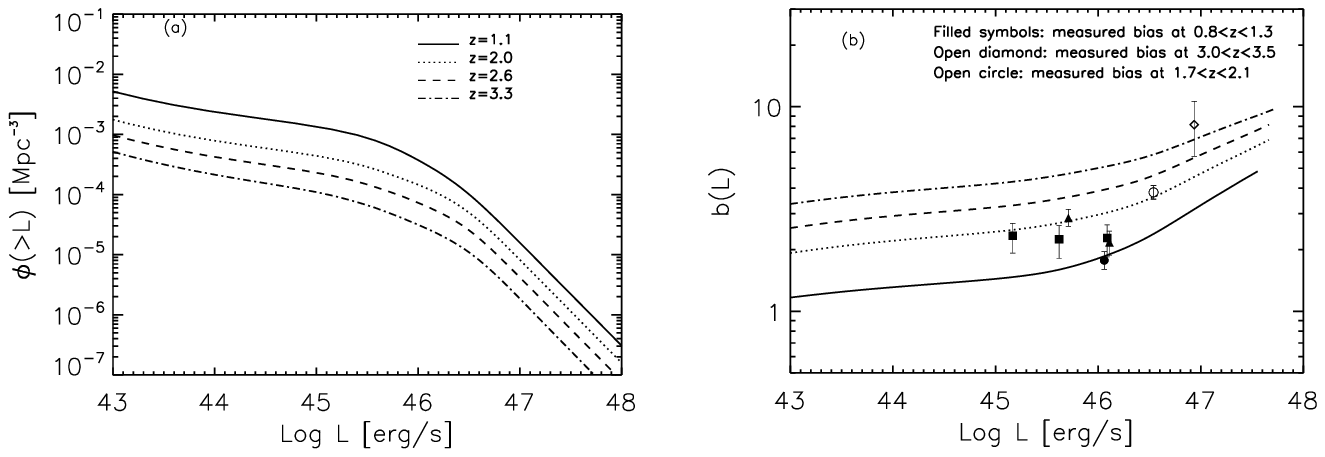}
\caption{Panel \emph{a}: cumulative space density of predicted hosts (including those
with inactive black holes of the same  mass) that have a luminosity
higher than $L$ for four different redshifts. Panel \emph{b}:
predicted bias as a function of luminosity and redshift,
computed according to equations~(\ref{eq|nhosts}) and
~(\ref{eq|bias}); the data are from da Angela et al. (2006; \emph{filled triangles}),
Myers et al. (2007; \emph{filled squares}), Porciani \& Norberg (2007; \emph{filled} and \emph{open circles}), and
Shen et al. (2007a; \emph{open diamonds}), all corrected to $\sigma_8=0.8$. Both panels refer to the predictions of our reference model.}
\label{fig|hosts}
\end{figure}

Observational estimates of the local black hole mass function still show substantial discrepancies
among different authors, depending on the correlation used to derive it (e.g., $M_{\bullet}-\sigma$, $M_{\bullet}-L_{\rm bulge}$,
$M_{\bullet}-M_{\rm star}$, $M_{\bullet}$-S\'{e}rsic index, or fundamental plane),
the calibration of the correlation, and the intrinsic scatter of the correlation.
Above $M_{\bullet}\sim 10^9\, {\rm M_{\odot}}$, estimates depend on extrapolation
of the observed scaling relations, and below $M_{\bullet}\sim 10^{7.5}\, {\rm M_{\odot}}$
they are sensitive to the treatment of spiral bulges. The grey band
in Figure~\ref{fig|LocalMFs} (and subsequent figures) encompasses most
estimates, but the fundamental plane (Hopkins et al. 2007b) and S\'{e}rsic
index (Graham et al. 2007) methods imply more sharply peaked mass functions. The integrated
mass densities of all of these estimates are in the range $\rho_{\bullet}\sim 3-5.5\times 10^5\, {\rm M_{\odot}\,Mpc^{-3}}$.

Our simplest models assume a single characteristic Eddington accretion rate $\dot{m}_0$,
independent of redshift and black hole mass, and all of our models assume a single
value of the radiative efficiency $\epsilon$. Matching the local
black hole mass density requires $\epsilon=0.075\times(\rho_{\bullet}/4.5\times 10^5\, {\rm M_{\odot}\,Mpc^{-3}})^{-1}$
for our standard estimate of the AGN LF, or $\epsilon \sim 0.094(\rho_{\bullet}/4.5\times 10^5\, {\rm M_{\odot}\,Mpc^{-3}})^{-1}$
for the HRH07 luminosity function.\footnote{More precisely, it is $\epsilon/(1-\epsilon)$ that
is proportional to $\rho_{\bullet}^{-1}$, but the difference from
$\epsilon\propto \rho_{\bullet}^{-1}$ is tiny over the allowed range.}
With $\epsilon$ thus fixed, the value of $\dot{m}_0$ determines
the peak of the predicted local black hole mass function in the
$M_{\bullet}\Phi(M_{\bullet})$ vs $M_{\bullet}$ plane. Note that
with our definitions the Eddington luminosity ratio
is $\lambda=L/L_{\rm Edd}\approx \dot{m}_0(\epsilon/0.1)$ (equation~\ref{eq|Lambda}).
Matching the observed
peak at $\log M_{\bullet}/{\rm M_{\odot}}\sim 8.5$
implies $\dot{m}_0\approx 0.6$ ($\lambda \approx 0.45$) for our standard LF estimate and
$\dot{m}_0\approx 1$ ($\lambda \approx 0.95$)
for the HRH07 LF. Lower values, $\dot{m}_0=0.1-0.3$, shift
the peak location to untenably high masses, $\log M_{\bullet}/{\rm M_{\odot}}\sim 8.9-9.3$
or, for HRH07, $\log M_{\bullet}/{\rm M_{\odot}}\sim 9.1-9.6$.
The single-$\dot{m}_0$ models achieve a reasonable match to our ``grey-band''
observational estimates of the width and asymmetry of the local mass function, though
for our LF estimate the predicted mass function is too high at $M_{\bullet}>10^9\, {\rm M_{\odot}}$
and is therefore somewhat too broad. Single-$\dot{m}_0$ models cannot
reproduce the more sharply peaked local mass functions estimated by Graham et al. (2007)
or Hopkins et al. (2007b).

Our reference model, which has $\epsilon=0.065$, $\dot{m}_0=0.60$,
and our standard LF estimate, predicts a duty cycle for activity of
$10^9\, {\rm M_{\odot}}$ black holes that declines steadily from
0.15 at $z=4$ to 0.07 ($z=3$), 0.035 ($z=2$), 0.004 ($z=1$), and
$10^{-4}$ ($z=0$). The decline in duty cycle for lower mass black
holes is much shallower. Massive black holes therefore build their
mass relatively early while low mass black holes grow later, the
phenomenon often referred to as ``downsizing''. Our results on mean
radiative efficiency and duty cycle evolution are also in
qualitative agreement with those found by Haiman et al. (2004). The
predicted duty cycles seem in reasonable accord with observational
estimates, though these estimates have considerable uncertainty and
do not, as yet, probe mass and redshift dependence in much detail.
The electronic tables described in \S~\ref{subsec|Tables} provide
tabulations as a function of redshift of our AGN bolometric LF
estimate, the HRH07 LF, and black hole mass functions and duty
cycles for single-$\dot{m}_0$ models that are in good agreement with
the observed $z=0$ mass functions given these LF inputs.

We have examined models in which the Eddington ratio accretion rate
$\dot{m}_0$ is reduced at low redshift or at low black hole mass.
Declining redshift evolution of $\dot{m}_0$ damps ``downsizing,''
reducing the dependence of duty cycles on black hole mass and
redshift. This model produces a typical duty cycle $P_0\sim
10^{-2.5}$ at $z=0$, about two orders of magnitude higher than in
the reference model and consistent with the local duty cycle
inferred from observations by Greene \& Ho (2007). In general terms,
the observed luminosity-dependent density evolution of the AGN LF
can be explained by preferential suppression of activity in high
mass black holes at low redshift, by a decline in the typical
accretion rate at low redshift, or by some combination thereof. The
mass-dependent $\dot{m}_0$ model associates more of the AGN
emissivity to high mass black holes, so it predicts a $z=0$ mass
function that is more sharply peaked, in better agreement with the
estimates of Hopkins et al. (2007b) and Graham et al. (2007) but
worse agreement with other estimates. This model predicts a stronger
mass dependence of duty cycles than our reference model because it
maps low mass black holes, whose abundance is already suppressed, to
less luminous, and hence more common, AGN.

We have also considered a model in which each black hole has a $50\%$ probability per Hubble time of merging
with another black hole of equal mass. Merger-driven growth in this
model has little impact on the black hole mass function until $z<1$, when
accretion-driven evolution has slowed. Low redshift mergers slightly
depress the low mass end of the $z=0$ mass function and significantly
enhance the high mass tail, worsening the agreement with observational
estimates. Models incorporating theoretically predicted merger rates can allow more realistic
calculations of the impact of mergers on the black hole population; this impact will
probably be smaller than in the simplified model considered here.

We have calculated the clustering bias of AGN as a function of luminosity and redshift
for our reference model, assuming a monotonic relation between black hole mass
and halo mass. The predictions are in reasonable accord with observational
estimates. We will examine AGN clustering predictions in more detail
in future work, with attention to what models can be excluded by the data and what can
be learned by matching the full AGN correlation function
in addition to an overall bias factor.

MHD simulations (e.g., Gammie et al. 2004; Shapiro 2005) show that disk accretion
onto Kerr black holes spins them up to an equilibrium spin rate $a\approx 0.95$
(where $a=1$ is the angular momentum parameter for a maximally rotating black hole). The radiative
efficiency in these models is $\epsilon\approx 0.16-0.2$.
These high efficiencies would lead to black hole mass densities a factor of two or more
below our central estimate, and below our estimated lower bound.
Furthermore, our results show that models with $\epsilon$ in the range
$0.06-0.11$ can achieve a good match to the overall shape of the
black hole mass function near the peak in $M_{\bullet}\Phi(M_{\bullet})$,
not just the value of $\rho_{\bullet}$, given
plausible choices of $\dot{m}_0$. Systematic uncertainties
in the AGN LF do not appear large enough to accommodate
$\epsilon\gtrsim 0.15$. Accommodating these high efficiencies
would instead require a substantial downward revision of recent
estimates of the local black hole mass function, reducing the integrated mass density
to $\rho_{\bullet}\sim 2\times 10^5\, {\rm M_{\odot}\,Mpc^{-3}}$. Our results
are consistent with a scenario like the one
of King \& Pringle (2006) in which chaotic accretion spins down
black holes because of counter-alignment with the accretion disk angular momentum,
or with other mechanisms that reduce efficiencies below
the MHD-simulation predictions.

The assumption that all active black holes at a given mass and
redshift have the same $\dot{m}_0$ is clearly an idealization, at
least in the local universe where observations indicate a wide range
of Eddington ratios (Heckman et al.\ 2005; Greene \& Ho 2007). Steed
\& Weinberg (2003) and Yu \& Lu (2004) discussed continuity equation
models evolved adopting a distribution of Eddington ratios. In
particular, Yu \& Lu (2004) have derived the relation between the
integrated number of AGNs shining at all times at a given luminosity
$L$, the mean light curve of black holes, and the local black hole
mass function. Following their equation (18), we find that a good
match between the cumulative number of AGNs and of relic black holes
can be achieved for $\epsilon\sim 0.07$ (required by the So{\l}tan
argument) and a mean AGN light curve exponentially increasing with
$\lambda=0.6$ and a negligible declining phase, similar to our
reference model. Alternatively, we find that a good match can be
obtained by assuming that black holes grow rapidly in a
Super-Eddington phase with $\lambda\gtrsim 2$ and then have a long
declining phase, qualitatively resembling our $\dot{m}(z)$ model
discussed in \S~\ref{subsec|etamdot}. In future work we will
investigate models that incorporate multiple $\dot{m}_0$-values and
accretion modes, including the addition of modest log-normal scatter
in $\dot{m}_0(M,z)$ (e.g., Kollmeier et al.\ 2005; Netzer et al.\
2007; Shen et al.\ 2007b) and sharper revisions in which some black
holes accrete at super-Eddington or highly sub-Eddington rates,
perhaps with reduced radiative efficiencies (Narayan, Mahadevan, \&
Quataert 1998 and references therein).  We will also incorporate
mergers at the rates predicted by theoretical models of cold dark
matter subhalos and their associated black holes (Yoo et al.\ 2007).
For appropriate parameter choices, we expect that many scenarios can
be made consistent with the observed AGN LF and the local black hole
mass function. However, direct measurements of Eddington ratio
distributions and measurements of AGN clustering and host
properties, all as a function of luminosity and redshift, should
greatly narrow the field of viable models.  Within the (often
substantial) uncertainties of existing data, a simple model in which
all black holes grow by accreting gas at mildly sub-Eddington rates
with a radiative efficiency $\epsilon \approx 0.06-0.1$ is
surprisingly successful at reproducing a wide range of observations.


\begin{acknowledgements}
This work was supported by NASA grant GRT000001640 and by
Spanish ministry grants AYA2006-06341 and AYA2006-15623-C02-01.
We thank
Alister Graham and Philip Hopkins for providing data on their
results on the local black hole mass function and John Silverman
for providing data on the X-ray luminosity function.
\end{acknowledgements}



\newpage


\begin{deluxetable}{cccccc}
\tablecolumns{6} \tablewidth{0pc} \tablecaption{AGN Luminosity
Function parameters} \tablehead{\colhead{Parameter} &  &  &  & & }
\startdata
$p_1$         &  $4.23$ ($z<4$) & $-0.615z+6.690$ & $3.0$\, ($z=6$)   &  &   \\
$p_2$         &  $-1.5$ at all $z$&   &  & & \\
$\gamma_1$     &  $0.86$ at all $z$&    &  & & \\
$\gamma_2$     &  $2.6$\, ($z<0.5$) & $-0.933z+3.067$ & $2.32$\, ($0.8<z<4$) & $0.24z+1.36$ & $2.8$\, ($z=6$)  \\
$A$           &  $5.04\times 10^{-6}\, {\rm Mpc^{-3}}$    &   &   & & \\
$z_c^*$       &  $1.9$ &     &  & & \\
$\log L_a$    &  $44.6$\, ($z<3$) & 0.0667$z$+44.4 & 44.8\, ($z=6$)    &   &   \\
$\log L_*$    &  $43.94$&     &  & & \\
\enddata
\tablecomments{ \, List of the parameters entering the AGN
luminosity function described in \S~\ref{sec|AGNLF}. Some of the
parameters assume different values in different redshift bins as
quoted. A smooth linear transition in redshift $z$ in the values has
been applied between discontinuous redshift bins. Luminosities are
the inferred values in the $2-10$ keV band, in units of ${\rm erg\,
s^{-1}}$.} \label{Table|LF}
\end{deluxetable}

\newpage

\begin{deluxetable}{cccccc}
\tablecolumns{6} \tablewidth{0pc} \tablecaption{AGN BOLOMETRIC
LUMINOSITY FUNCTIONS}
\tablehead{$z$ & $\log L$ & $\log \Phi$  & $\log \Phi_{\rm no CT}$ & $\log \Phi_{\rm Double CT}$ & $\log \Phi_{\rm HRH07}$\\
(1) & (2) & (3) & (4)  & (5) & (6)} \startdata
0.020 & 41.00 & -1.859 & -1.994 & -1.757 & -1.374\\
0.020 & 41.25 & -2.074 & -2.208 & -1.972 & -1.578\\
..........\\
0.020 & 48.00 & -10.08 & -10.16 & -10.01 & -9.539\\
0.260 & 41.00 & -1.903 & -2.036 & -1.796 & -1.609\\
0.260 & 41.25 & -2.089 & -2.222 & -1.984 & -1.787\\
0.260 & 41.50 & -2.267 & -2.399 & -2.162 & -1.967\\
..........\\
\enddata
\tablecomments{\,(1) redshift; (2) bolometric luminosity in
logarithmic scale in units of ${\rm erg\, s^{-1}}$; (3) reference
bolometric AGN luminosity function in logarithmic scale in units of
${\rm Mpc^{-3}\, dex^{-1}}$; (4) reference bolometric AGN luminosity
function with no Compton-thick sources (i. e., with $\log N_H/{\rm
cm^{-2}}>24$); (5) reference bolometric AGN luminosity function with
Double contribution from Compton-thick sources (i. e., with $\log
N_H/{\rm cm^{-2}}\le 26$); (6) bolometric AGN luminosity function
from HRH07. The full table is available in electronic form.}
\label{Table|FitsLF}
\end{deluxetable}

\newpage

\begin{deluxetable}{ccccccc}
\tablecolumns{7} \tablewidth{0pc} \tablecaption{BLACK HOLE MASS
FUNCTIONS} \tablehead{$z$ & $\log M_{\bullet}$ & $\log \Phi$  &
$\log \Phi_{\rm no CT}$ &
$\log \Phi_{\rm Double CT}$ & $\log \Phi_{\rm HRH07}$ & $\log \Phi_{\rm HRH07_{\rm EXTRA}}$\\
(1) & (2) & (3) & (4)  & (5) & (6) & (7)} \startdata
0.02 & 5.0 & -1.264 & -1.398 & -1.162 & -0.928 & -0.952\\
0.02 & 5.2 & -1.394 & -1.528 & -1.292 & -1.060 & -1.082\\
..........\\
0.02 & 9.6 & -4.683 & -4.779 & -4.604 & -4.436 & -4.964\\
0.26 & 5.0 & -1.482 & -1.614 & -1.376 & -1.129 & -1.131\\
0.26 & 5.2 & -1.608 & -1.739 & -1.501 & -1.247 & -1.248\\
0.26 & 5.4 & -1.723 & -1.855 & -1.616 & -1.365 & -1.364\\
..........\\
\enddata
\tablecomments{\,(1) redshift; (2) black hole mass in logarithmic
scale in units of ${\rm M_{\odot}}$; (3) black hole mass function in
our reference model in logarithmic scale in units of ${\rm
Mpc^{-3}\, dex^{-1}}$; (4) reference black hole mass function in the
same model but with no Compton-thick sources; (5) reference black
hole mass function in the same model but with Double contribution
from Compton-thick sources; (6) black hole mass function in the same
model but adopting the HRH07 luminosity function; (7) black hole
mass function obtained adopting the HRH07 luminosity function with
$\epsilon=0.09$ and $\dot{m}=1$, as in Figure~\ref{fig|Models}d. The
full table is available in electronic form.} \label{Table|Fits}
\end{deluxetable}
%
\begin{deluxetable}{ccccccc}
\tablecolumns{7} \tablewidth{0pc} \tablecaption{BLACK HOLE DUTY
CYCLES} \tablehead{$z$ & $\log M_{\bullet}$ & $P_0$  & $P_0^{\rm no
CT}$ &
$P_0^{\rm Double CT}$ & $P_0^{\rm HRH07}$ & $P_0^{\rm HRH07_{\rm EXTRA}}$\\
(1) & (2) & (3) & (4)  & (5) & (6) & (7)} \startdata
0.02 &   5.0 &   0.00995 &  0.00995 &   0.00995 &  0.01401  &  0.00727\\
0.02 &   5.2 &   0.00941 &  0.00942 &   0.00942 &  0.01299  &  0.00668\\
..........\\
0.02 &   9.6 &   0.00008 &  0.00008 &   0.00008 &  0.00011  &  0.00009\\

0.26 &   5.0 &   0.03063 &  0.03057 &   0.03050 &  0.01947  &  0.01050\\
0.26 &   5.2 &   0.03246 &  0.03231 &   0.03215 &  0.01837  &  0.00987\\
0.26 &   5.4 &   0.03273 &  0.03243 &   0.03212 &  0.01729  &  0.00924\\
..........\\
\enddata
\tablecomments{\,(1) redshift; (2) black hole mass in logarithmic
scale in units of ${\rm M_{\odot}}$; (3) duty cycle for our
reference model; (4) duty cycle for the reference model with no
Compton-thick sources; (5) duty cycle for the reference model with
Double contribution from Compton-thick sources; (6) duty cycle for
the model obtained adopting the HRH07 luminosity function; (7) duty
cycle for the model obtained adopting the HRH07 luminosity function
with $\epsilon=0.09$ and $\dot{m}=1$. The full table is available in
electronic form.} \label{Table|DutyCycles}
\end{deluxetable}

%
\newpage

\newpage
\begin{thebibliography}{}
\bibitem{} Alexander, D. M., et al. 2003, AJ, 126, 539
\bibitem{} Antonucci, R. R. 1993, ARA\&A, 31, 473
\bibitem{} Babic, A., Miller, L., Jarvis, M.J., Turner, T.J., Alexander, D.M., \& Croom, S.M. 2007, A\&A,
474, 755
\bibitem{} Ballantyne, D. R., \& Papovich, C. 2007, ApJ, 660, 988
\bibitem{} Barger, A. J., Cowie, L. L., Capak, P., Alexander, D. M.,
Bauer, F. E., Brandt, W. N., Garmire, G. P. \& Hornschemeier, A. E.
2003, ApJ, 584, L61
\bibitem{} Barger , A. J., \& Cowie, L. L. 2005, ApJ, 635, 115
\bibitem{} Barger, A. J., et al. 2005, AJ, 129, 578
\bibitem{} Batcheldor, D., Marconi, A., Merritt, D., \& Axon, D. J.
2007, ApJL, 663, 85
\bibitem{} Bell, E. F., Phleps, S., Somerville, R. S., Wolf, C., Borch, A., \& Meisenheimer, K. 2006, ApJ, 652, 270
\bibitem{} Bentz, M. C., Peterson, B. M., Pogge, R. W., Vestergaard, M., \& Onken, C.
A. 2006, ApJ, 644, 133
\bibitem{} Bernardi, M., Sheth, R. K., Tundo, E., \& Hyde, J. B. 2007, ApJ, 660, 267
\bibitem{} Binney, J., \& Merrifield, M. 2000, \emph{Galactic Astronomy},
Princeton Univ. Press
\bibitem{} Bolton, J. S., Haehnelt, M. G., Viel, M., Springel, V.
2005, MNRAS, 357, 1178
\bibitem{} Bongiorno, A., et al. 2007, A\&A, 472, 443
\bibitem{} Borys, C., Smail, I., Chapman, S. C., Blain, A. W.,
  Alexander, D. M., Ivison, R. J. 2005, ApJ, 635, 853
\bibitem{} Brown, M. J. I., et al. 2006, ApJ, 638, 88
\bibitem{} Bundy, K., et al. ApJ, submitted, arXiv/0710.2105
\bibitem{} Caputi, K. I., McLure, R. J., Dunlop, J. S., Cirasuolo,
M., \& Schael, A. M. 2006, MNRAS, 366, 609
\bibitem{} Cavaliere, A., Morrison, P.,
\& Wood, K. 1971, ApJ, 170, 223
\bibitem{} Cavaliere, A., \& Vittorini, V. 2000, ApJ,
543, 599
\bibitem{} Churazov, E., et al. 2007, A\&A, 467, 529
\bibitem{} Comastri, A., Setti, G., Zamorani, G., \& Hasinger, G.
1995, A\&A, 296, 1
\bibitem{} Comastri, A. 2004, Review for
{\it "Supermassive Black Holes in the Distant Universe"}, Ed. A. J.
Barger, Kluwer Academic, astroph/0403693
\bibitem{} Conroy, C., Ho, S., \& White, M. 2007, MNRAS, 379, 1491
\bibitem{} Constantin, A., \& Vogeley, M. S. 2006, 650, 727
\bibitem{} Cool, R. J., Kochanek, C. S., Eisenstein, D. J., Stern, D.,
Brand, K., Brown, M. J. I., Dey, A., Eisenhardt, P. R., Fan, X.,
Gonzalez, A. H., Green, R. F., Jannuzi, B. T., McKenzie, E. H.,
Rieke, G. H., Rieke, M., Soifer, B. T., Spinrad, H., \& Elston, R.
J. 2006, AJ, 132, 823
\bibitem{} Cowie, L. L., et al. 2002, ApJ, 566, L5
\bibitem{} Cowie, L. L., et al. 2003, ApJ, 584, L57
\bibitem{} Croom, S. M., et al. 2004, MNRAS, 349, 1397
\bibitem{} Croton, D. J., Springel, V., White, S. D. M., De Lucia,
G., Frenk, C. S., Gao, L., Jenkins, A., Kauffmann, G., Navarro, J.
F., \& Yoshida, N. 2006, MNRAS, 365, 11
\bibitem{} Croton, D. J. 2006b, MNRAS, 369, 1808
\bibitem{} da Angela, J., et al. 2006, MNRAS, submitted,
astroph/0612401
\bibitem{} Dasyra, K. M., Tacconi, L. J., Davies, R. I., Naab, T., Genzel,
R., Lutz, D., Sturm, E., Baker, A. J., Veilleux, S., Sanders, D. B.,
\& Burkert, A. 2006, ApJ, 651, 835
\bibitem{} Dasyra, K. M., Tacconi, L. J., Davies, R. I., Genzel, R., Lutz,
D., Peterson, B. M., Veilleux, S., Baker, A. J., Schweitzer, M., \&
Sturm, E. 2007, ApJ, 657, 102
\bibitem{} De Zotti, G., Shankar, F., Lapi, A., Granato, G. L.,
Silva, L., Cirasuolo, M., Salucci, P., \& Danese, L. 2006, MmSAIt,
77, 661
\bibitem{} Elvis, M., Risaliti, G., \& Zamorani, G. 2002, ApJ, 565,
L75
\bibitem{} Fabian, A. C., \& Iwasawa, K. 1999, MNRAS, 303, 34
\bibitem{} Fan, X., et al. 2001, AJ, 121, 54
\bibitem{} Fan, X., et al. 2004, ApJ, 128, 515
\bibitem{} Fardal, M. A., Katz, N., Weinberg, D. H., \& Dav\'e, R. 2007,
MNRAS, 379, 985
\bibitem{} Ferrarese, L., \& Merritt, D. 2000, ApJ, 539, L9
\bibitem{} Ferrarese, L. 2002, Proceedings of the 2nd KIAS Astrophysics Workshop, Seoul, Korea, astroph/0203047
\bibitem{} Fioc, M., \& Rocca-Volmerange, B. 1997, A\&A, 326, 950
\bibitem{} Fiore, F., et al. 2003, A\&A, 409, 79
\bibitem{} Fontana, A., et al. 2006, A\&A, 459, 745
\bibitem{} Fontanot, F., Cristiani, S., Monaco, P., Nonino, M.,
Vanzella, E., Brandt, W. N., Grazian, A., \& Mao, J. 2007, A\&A,
461, 39
\bibitem{} Fontanot, F., Monaco, P., Cristiani, S., \& Tozzi, P.
2006, MNRAS, 373, 1173
\bibitem{} Frontera, F., Orlandini, M.,
Landi, R., Comastri, A., Fiore, F., Setti, G., Amati, L., Costa, E.,
Masetti, N., \& Palazzi, E. 2007, ApJ, 666, 86
\bibitem{} Fukugita, M., Hogan, C. J., \& Peebles, P. J. E. 1998, ApJ, 503,
518
\bibitem{} Gammie, C. F., Shapiro, S. L., \& McKinney, J, C. 2004,
ApJ, 602, 312
\bibitem{} Gilli, R., Comastri, A., \& Hasinger, G. 2006, A\&A, 463,
79
\bibitem{} Graham, A. W., Erwin, P., Caon, N., \& Trujillo, I. 2001, ApJ, 563,
L11
\bibitem{} Graham, A. W., et al. 2007, MNRAS, 378, 198
\bibitem{} Graham, A. W. 2007, MNRAS, 379, 711
\bibitem{} Granato, G. L., De Zotti, G., Silva, L., Bressan, A., \&
 Danese, L. 2004, ApJ, 600, 580
\bibitem{} Granato, G. L., Silva, L., Lapi, A., Shankar, F., De Zotti, G., Danese,
L. 2006, MNRAS, 368L, 72
\bibitem{} Greene, J. E., \& Ho, L. C. 2006, ApJ, 641, 21
\bibitem{} Greene, J. E., \& Ho, L. C. 2007, ApJ, 667, 131
\bibitem{} Greve, T. R., et al. 2005, MNRAS, 359, 1165
\bibitem{} Kennefick, J. D., Djorgovski, S. G., \& de Carvalho, R. R.
1995, AJ, 110, 2553
\bibitem{} Haiman, Z., Ciotti, L., Ostriker, J. P. 2004, ApJ, 606, 763
\bibitem{} H\"{a}ring, N., \&  Rix, H. W.  2004, ApJ, 604, 89
\bibitem{} Hasinger, G., et al. 2001, A\&A, 365, 45
\bibitem{} Hasinger, G., Miyaji, T., Schmidt, M. 2005, A\&A, 441,
417
\bibitem{} Heckman, T. M., Kauffmann, G., Brinchmann, J.,
Charlot, S., Tremonti, C., \& White, S. D. M. 2004, ApJ, 613, 109
\bibitem{} Hickox, R. C. et al.\ 2007, ApJ, in press, astroph/0708.3678
\bibitem{} Hopkins, A. W., \& Beacom, J. F. 2006, 651, 142
\bibitem{} Hopkins, P. F., Hernquist, L., Cox,
T. J., Di Matteo, T., Robertson, B., \& Springel, V. 2006a, ApJS,
163, 1
\bibitem{} Hopkins, P. F., Robertson, B., Krause, E., Hernquist,
L., \& Cox, T. J. 2006b, ApJ, 652, 107
\bibitem{} Hopkins, P. F., Richards, G. T., Hernquist, L.
2007a, ApJ, 654, 731 (HRH07)
\bibitem{} Hopkins, P. F., Hernquist, L., Cox, T. J., Robertson, B., \& Krause,
E. 2007b, ApJ, in press, astroph/0701351
\bibitem{} Hopkins, P. F., Lidz, A., Hernquist, L., Coil, A. L., Myers, A. D., Cox, T. J., \& Spergel,
D. 2007c, ApJ, 662, 110
\bibitem{} Hosokawa, T. 2004, ApJ, 606, 139
\bibitem{} Hughes, S. A., \& Blandford, R. D. 2003, ApJ, 585, L10
\bibitem{} Hunt, M. P., Steidel, C. C.,
Adelberger, K. L., \& Shapley, A. E. 2004, ApJ, 605, 625
\bibitem{} Islam, R. R., Taylor, J. E., \& Silk, J. 2004, MNRAS,
354, 427
\bibitem{} Jiang, L., et al. 2006, AJ, 131, 2788
\bibitem{} Kaspi, S., et al. 2000, ApJ, 533, 631
\bibitem{} Kembhavi, A. K., \& Narlikar, J. V. 1999, in ``Quasars and Active Galactic
Nuclei", Cambridge Univ. Press, Cambridge
\bibitem{} King, A. R., \& Pringle, J. E. 2006, MNRAS, 373, 90
\bibitem{} Kollmeier, J. A., et al. 2006, ApJ, 648, 128
\bibitem{} La Franca, F., et al. 2005, ApJ, 635, 864
\bibitem{} Lapi, A., Shankar, F., Mao, J., Granato, G. L.,
Silva, L., De Zotti, G., \& Danese, L. 2006, ApJ, 650, 42
\bibitem{} Lauer, T. R., Faber, S. M., Richstone, D., Gebhardt, K.,
 Tremaine, S., Postman, M., Dressler, A., Aller, M. C., Filippenko, A. V., Green, R., Ho, L. C., Kormendy, J., Magorrian, J., \& Pinkney,
 J. 2006, ApJ, 670, 249
\bibitem{} Lauer, T. R., et al. 2007a, ApJ, 662, 808
\bibitem{} Lauer, T. R., Tremaine, S., Richstone, D., \& Faber, S.
M., 2007b, ApJ, 670, 249
\bibitem{} Lawrence, J. K. 1991, MNRAS, 252, 586
\bibitem{} Lidz, A., Hopkins, P. F., Cox, T. J., Hernquist, L., \&
Robertson, B. 2006, ApJ, 641, 41
\bibitem{} Mahabal, A., Stern, D., Bogosavljevic, M., Djorgovski,
S.G., \& Thompson, D. 2005, ApJ, 634, L9
\bibitem{} Magorrian, J., et al. 1998, AJ, 115, 2285
\bibitem{} Magdziarz, P., \& Zdziarski, A. A. 1995, MNRAS, 273, 837
\bibitem{} Malbon, R. K., Baugh, C. M., Frenk, C. S., \& Lacey, C.
G. 2006, MNRAS, submitted, astroph/0607424
\bibitem{} Maller, A. H., Katz, N., Kere\v{s}, D., Dav\'{e}, R., \& Weinberg, D. H. 2006, ApJ, 647, 763
\bibitem{} Marconi, A., \& Hunt, L. 2003, ApJL, 589, L21
\bibitem{} Marconi, A., Risaliti, G., Gilli, R., Hunt, L. K.,
Maiolino, R., \& Salvati, M. 2004, MNRAS, 351, 169
\bibitem{} Marshall, F. E., et al. 1980, ApJ, 235, 4
\bibitem{} Martinez-Sansigre, A. et al., 2005, Nature, 436, 666
\bibitem{} Masjedi, M., Hogg, D. W., \& Blanton, M. R. 2008, ApJ,
679, 260
\bibitem{} Matt, G., Fabian, A. C., Guainazzi, M., Iwasawa, K., Bassani, L., \& Malaguti,
G. 2000, MNRAS, 318, 173
\bibitem{} McLure, R. J., Dunlop., J. S. 2002, MNRAS, 331, 795
\bibitem{} McLure, R. J., Dunlop., J. S. 2004, MNRAS, 352, 1390
\bibitem{} McLure, R. J., Jarvis, M. J.,
Targett, T. A., Dunlop, J. S., \& Best, P. N. 2006, MNRAS, 368, 1359
\bibitem{} Merloni, A. 2004, MNRAS, 353, 1035
\bibitem{} Merloni, A., Rudnick, G., \& Di Matteo, T. 2004, MNRAS,
354, 37
\bibitem{} Merritt, D., \& Milosavljevi\'{c} M. 2005, LRR, 8, 8
\bibitem{} Miyaji, T., Hasinger, G., \& Schmidt, M. 2000, A\&A, 353, 25
\bibitem{} Myers, A. D., Brunner, R. J., Nichol, R. C., Richards, G. T., Schneider, D. P., \& Bahcall, N.
A. 2007, ApJ, 658, 85
\bibitem{} Miralda-Escud\'{e}, J. 2003, ApJ, 597, 66
\bibitem{} Monaco, P., \& Fontanot, F. 2005, MNRAS, 359, 283
\bibitem{} Murray, N., Quataert, E., Thompson, T. A. 2005, ApJ, 618, 569
\bibitem{} Nakamura, O., Fukugita, M., Yasuda, N., Loveday, J.,
Brinkmann, J., Schneider, D. P., Shimasaku, K., \& SubbaRao, M.
2003, AJ, 125, 1682
\bibitem{} Nandra, K., Mushotzky, R. F., Arnaud, K., Steidel, C. C.,
Adelberger, K. L., Gardner, J. P., Teplitz, H. I., \& Windhorst, R.
A. 2002, ApJ, 576, 625
\bibitem{} Nandra, K., Laird, E. S., \& Steidel, C. C. 2005, MNRAS,
360, L39
\bibitem{} Narayan, R., Mahadevan, R., \& Quataert, E. 1998,
in {\it "The Theory of Black Hole Accretion Discs"}, eds. M. A.
Abramowicz, G. Bjornsson, and J. E. Pringle (Cambridge: Cambridge
Univ. Press), p.148, astroph/9803141
\bibitem{} Netzer, H., Trakhtenbrot, B. 2007, ApJ, 654, 754
\bibitem{} Netzer, H., Lira, P., Trakhtenbrot, B., Shemmer, O., \& Cury, I. 2007, ApJ,
671, 1256
\bibitem{} Osmer, P. S. 2004, in Carnegie Observatories Astrophysics Series,
Vol. 1: Coevolution of Black Holes and Galaxies, ed. L.C. Ho
(Cambridge Univ. Press), in press, astroph/0304150
\bibitem{} Pannella, M., Hopp, U., Saglia, R. P., Bender, R.,
Drory, N., Salvato, M., Gabasch, A., \& Feulner, G. 2006, ApJ, 639,
1
\bibitem{} Peeples, M. S., \& Martini, P. 2006, ApJ, 652, 1097
\bibitem{} Pei, Y. C. 1995, ApJ, 438, 623
\bibitem{} Peng, C. Y., Impey, C. D., Rix, H. W., Kochanek, C. S.,
Keeton, C. R., Falco, E. E., Leh\'{a}r, J., \& McLeod, B. A. 2006,
ApJ, 649, 616
\bibitem{} Polletta, M., Weedman, D., Hoenig, S., Lonsdale, C. J.,
  Smith, H. E., \& Houck, J. \ 2008, ApJ, 675, 960
\bibitem{} Porciani, C., Magliocchetti, M., \& Norberg, P. 2004
MNRAS, 355, 1010
\bibitem{} Porciani, C., \& Norberg, P. 2006, MNRAS, 371, 1824
\bibitem{} Rees, M. J. 1984, ARA\&A, 22, 471
\bibitem{} Richards, G. T., et al. 2006, AJ, 131, 2766
\bibitem{} Richstone, D., et al. 1998, Nat, 395, A14
\bibitem{} Rigby, J. R., Rieke, G. H., Donley, J. L.,
Alonso-Herrero, A., \& P\'{e}rez-Gonz\'{a}lez, P. G. 2006, ApJ, 645,
115
\bibitem{} Salucci, P., Szuszkiewicz, E., Monaco, P., \& Danese, L.
1999, MNRAS, 307, 637
\bibitem{} Salpeter, E. E. 1964, ApJ, 140, 796
\bibitem{} Sanders, D. B., et al. 1988, ApJ, 325, 74
\bibitem{} Schirber, M., \& Bullock, J. S. 2003, ApJ, 584, 110
\bibitem{} Shankar, F., Salucci, P., Granato, G. L., De Zotti, G., \& Danese,
L. 2004, MNRAS, 354, 1020 (S04)
\bibitem{} Shankar, F., Lapi, A., Salucci, P., De Zotti, G., \& Danese,
L. 2006, ApJ, 643, 14
\bibitem{} Shankar, F., \& Mathur, S. 2007, ApJ, 660, 1051
\bibitem{} Shankar, F., \& Ferrarese, L. 2008, ApJ, submitted
\bibitem{} Shankar, F., Bernardi, M., \& Z\'{o}ltan, H. 2008, ApJ,
submitted
\bibitem{} Shapiro, S. L. 2005, ApJ, 620, 59
\bibitem{} Shen, Y., et al. 2007a, AJ, 133, 2222
\bibitem{} Shen, Y., Greene, J. E., Strauss, M., Richards, G. T., \& Schneider, D. P. 2007b, ApJ, submitted, arXiv/0709.3098
\bibitem{} Sheth, R. K., \& Tormen, G. 1999, MNRAS, 308, 119
\bibitem{} Sheth, R. K., Hui, L., Diaferio, A., \& Scoccimarro, R.
2001, MNRAS, 325, 1288
\bibitem{} Sheth, R. K., et al. 2003, ApJ, 594, 225
\bibitem{} Shields, G. A., Menezes, K. L., Massart, C. A., \& Vanden Bout,
P. 2006, ApJ, 641, 683
\bibitem{} Shinozaki, K., Miyaji, T., Ishisaki, Y., Ueda, Y., \& Ogasaka,
Y. 2006, 131, 2843
\bibitem{} Silverman, J. D., et al. 2008, ApJ, 679, 118
\bibitem{} Small, T. A., \& Blandford, R. D. 1992, MNRAS, 259, 725
\bibitem{} Smith, R. E., Peacock, J. A., Jenkins, A., White, S. D. M.,
Frenk, C. S., Pearce, F. R., Thomas, P. A., Efstathiou, G., \&
Couchman, H. M. P. 2003, MNRAS, 341, 1311
\bibitem{} So{\l}tan, A. 1982, MNRAS, 200, 115
\bibitem{} Steed, A., \& Weinberg, D. H. 2003, astroph/0311312 (SW03)
\bibitem{} Steffen, A. T., Strateva, I., Brandt, W. N., Alexander,
D. M., Koekemoer, A. M., Lehmer, B. D., Schneider, D. P, \& Vignali,
C. 2006, AJ, 131, 2826
\bibitem{} Steidel, C. C., Hunt, M. P., Shapley, A. E.,
Adelberger, K. L., Pettini, M., Dickinson, M., \& Giavalisco, M.
2002, ApJ, 576, 653
\bibitem{} Stern, D., et al. 2005, ApJ, 631, 163
\bibitem{} Tacconi, L. J., Neri, R., Chapman, S. C., Genzel, R., Smail,
I., Ivison, R. J., Bertoldi, F., Blain, A., Cox, P., Greve, T.,
Omont, A. 2006, 640, 228
\bibitem{} Tamura, N., Ohta, K., \& Ueda, Y. 2006, MNRAS, 365, 134
\bibitem{} Tinker, J. L., Weinger, D. H., Zheng Z., \& Zehavi, I.
2005, ApJ, 631, 41
\bibitem{} Tonry, J., et al. 2001, ApJ, 546, 681
\bibitem{} Tozzi, P., et al. 2006, A\&A, 451, 457
\bibitem{} Treister, E., et al. 2006, ApJ, 640, 603
\bibitem{} Treu, T., Ellis, R. S., Liao, T. X., van Dokkum, P. G.,
Tozzi, P., Coil, A., Newman, J., Cooper, M. C., \& Davis, Marc 2005,
ApJ, 633, 174
\bibitem{} Tundo, E., Bernardi, M., Hyde, J. B., Sheth, R. K., \& Pizzella,
A. 2007, ApJ, 663, 53
\bibitem{} Ueda, Y., Akiyama, M., Ohta, K., \& Miyaji, T. 2003, ApJ, 598, 886 (U03)
\bibitem{} Vale, A.,\& Ostriker, J. P. 2004, MNRAS, 353, 189
\bibitem{} Vestergaard, M. 2004, ApJ, 601, 676
\bibitem{} Vittorini, V., Shankar, F., \& Cavaliere, A. 2005, MNRAS,
363, 1376
\bibitem{} Volonteri, M., Madau, P., \& Haardt, F. 2003, ApJ, 593,
661
\bibitem{} Volonteri, M., Madau, P., Quataert, E., \& Rees, M. J. 2005,
ApJ, 620, 69
\bibitem{} Volonteri, M., Salvaterra, R., \& Haardt, F. 2006, MNRAS,
373, 121
\bibitem{} Wilman, R. J., \& Fabian, A. C. 1999, MNRAS 309, 862
\bibitem{} Wisotzki, L. 1999, Reviews in Modern Astronomy 12,
Astronomical Instruments and Methods at the turn of the 21st
Century. Edited by Reinhard E. Schielicke. Published by
Astronomische Gesellschaft, Hamburg, p. 231
\bibitem{} Wolf, C., et al. 2003, A\&A, 408, 499
\bibitem{} Wyithe, J. S. B. 2004, MNRAS, 1082, 1098
\bibitem{} Yoo, J., \& Miralda-Escud\'{e}, J. 2004, ApJ, 614, 25
\bibitem{} Yoo, J., Miralda-Escud\'{e}, J., Weinberg, D. H., Zheng, Z., \& Morgan,
C. W. 2007, ApJ, submitted, astroph/0702199
\bibitem{} Yu, Q., \& Lu, Y. 2004, ApJ, 602, 603
\bibitem{} Yu, Q., \& Tremaine, S. 2002, MNRAS, 335, 965
\end{thebibliography}
\end{document}